\documentclass{article}
\usepackage{framed,multirow}

\usepackage{amssymb}
\usepackage{amsmath}
\usepackage{latexsym}
\usepackage{subcaption}
\usepackage[section]{placeins}
\usepackage{hyperref}
\usepackage[margin=1in]{geometry}
\usepackage{authblk}
\usepackage{hyperref}
\usepackage[british]{babel}
\usepackage{graphicx}

\usepackage{url}
\usepackage{xcolor}
\definecolor{newcolor}{rgb}{.8,.349,.1}

\begin{document}

\title{Travel times and ray paths for acoustic and elastic waves in generally anisotropic media}

\author[1]{James Ludlam \footnote{email: james.ludlam@strath.ac.uk (James Ludlam)}}

\author[1]{Katherine Tant}
\author[1]{Victorita Dolean}
\author[2]{Andrew Curtis}
\affil[1]{University of Strathclyde, 26 Richmond St, Glasgow G1 1XQ, UK}
\affil[2]{University of Edinburgh, School of Geosciences, Edinburgh EH9 3FE, UK}

\maketitle

\begin{abstract}
Wavefield travel time tomography is used for a variety of purposes in acoustics, geophysics and non-destructive testing. Since the problem is non-linear, assessing uncertainty in the results requires many forward evaluations. It is therefore important that the forward evaluation of travel times and ray paths is efficient, which is challenging in generally anisotropic media. Given a computed travel time field, ray tracing can be performed to obtain the fastest ray path from any point in the medium to the source of the travel time field. These rays can then be used to speed up gradient based inversion methods. We present a forward modeller for calculating travel time fields by localised estimation of wavefronts, and a novel approach to ray tracing through those travel time fields. These methods have been tested in a complex anisotropic weld and give travel times comparable to those obtained using finite element modelling while being computationally cheaper.
\end{abstract}

\section{Introduction}
Ultrasonic non-destructive evaluation (NDE) is an umbrella term that describes a range of techniques used to check for flaws during the manufacture and maintenance of safety-critical infrastructure built from materials such as metals and concrete \cite{schabowicz_ultrasonic_2014}. Usually this involves imaging the medium using algorithms that embed the assumption that the materials are homogeneous and isotropic \cite{connolly_application_2009}. However, for many materials of interest this is not the case. For example, in common polycrystalline welds, there exists a locally anisotropic grain structure which causes waves to scatter and refract. Using homogeneous assumptions in this case can dramatically reduce the reliability of the detection and characterisation of material defects. Given correct information about the spatially varying material properties, the heterogeneity and anisotropy can be compensated for to improve the accuracy of estimated defect properties \cite{connolly_correction_2010}.\par

In geophysical travel time tomography of the solid Earth's structure, orientations and properties of rocks and of the fluid reservoirs that they contain are imaged using travel times of seismic or electromagnetic waves measured between wave energy sources and receivers. Almost all rocks are significantly heterogeneous and anisotropic, and such effects must be accounted for and indeed imaged for purposes ranging from assessing earthquake hazards to monitoring subsurface dynamics \cite{LIU201229}. In both NDE and geophysics, it is therefore necessary to model the effects of heterogeneous, anisotropic materials on waves.
\par

Traveltime tomography is a procedure that allows us to use observed travel times (the fastest time for waves to travel between two known points on or in the object or medium) to estimate the spatially varying material properties within that object by comparing observed travel times with modelled time of flight data \cite{marioli_digital_1992}. Traveltime tomography is often preferred over full waveform techniques because calculating travel times is computationally cheaper than modelling the full scattered wave field \cite{virieux_overview_2009}. Due to the complexity of the travel time tomography problem, Monte Carlo sampling methods are often deployed to find the solution and assess its uncertainty, but these methods are computationally expensive for large datasets and high dimensional parametrised systems \cite{bodin_sambridge, galetti_2015, haario_markov_2004}. Variational Bayesian Inversion (VBI) approaches can offer a computationally efficient approach for these inverse problems by formulating them as an optimisation problem instead of a sampling problem \cite{zhang_seismic_2020, keng_variational_2017, seeger_variational_2010}. Nevertheless, for many tomographic inversion techniques wave field data have to be modelled many times so computational efficiency is key.

We would like to image the interior of anisotropic materials, and in particular to invert travel time data collected on the surface of an object for grain orientation maps. In order to achieve this we need a forward model that calculates travel times of first arriving energy travelling between source and receiver locations. We also want corresponding ray paths along which the energy travelled. These ray paths can be used to increase the efficiency of inverse methods that require gradients of the travel times with respect to parameters that represent properties of the medium. Travel time fields can be used to find the fastest ray path between two given points in some medium. Fast Marching Methods (FMM) \cite{sethian_fast_1999, tant_effective_2020, sethian1999_3D, Alkhalifah2001} and Fast Sweeping Methods \cite{zhao2005fast, Lan2018, Umair2015, Le_Bouteiller2017, desquilbet_single_2021} are two such efficient methods for calculating travel time fields which have similar performance \cite{FMM_FSM_comp}. In this work we employ a Fast Marching Method and while the FMM has been used to calculate travel time fields in isotropic media \cite{sethian_fast_1999, sethian1999_3D}, an extension is required for anisotropic media. Typically, previous extensions of the FMM to anisotropic media have limited to weak, elliptical or tilted transversely isotropic media \cite{sethian2004, Yingyu_2022}. For generally anisotropic media, the Anisotropic Multi-stencil Fast Marching Method (AMSFMM) \cite{tant_effective_2020} can be used, where materials are defined using vectors describing the group and phase velocity curves, facilitating the study of any material with smooth velocity curves and $90^\circ$ rotational symetry. However the AMSFMM is unsuitable for ray tracing as its travel time errors are directionally biased. Note that although ray tracing can be performed in isotropic media using gradient descent through the travel time field (since the ray path is perpendicular to the wavefront), this is not the case for anisotropic media and so a different ray tracing method is required.\par
In this paper we present a novel forward modelling algorithm for calculating travel time fields, the Anisotropic Locally Interpolated Fast Marching Method (ALI-FMM), and a novel ray tracing method which uses ALI-FMM to calculate the fastest ray path in anisotropic media. Specifically, we study the arrival times of longitudinal waves in two dimensional orthotropic media. However, the framework retains the flexibility of the AMSFMM to facilitate the study of generally anisotropic media restricted only by smoothness and the presence of $180^{\circ}$ rotational symmetry. The success of the methodology is examined in the context of reconstructing a complex and strongly anisotropic velocity structure derived from a real weld. This represents an example typical of non-destructive testing applications. These methods are thus shown to enable travel time tomography to be applied in highly heterogeneous and anisotropic media. 

\section{Fast Marching Method in Anisotropic Media}
A travel time field gives the shortest travel time from a fixed source point to all other points in the domain $I = X \times Y = \{ (x_i, y_j) |x_i \in X, y_j \in Y \}$ where $X$ and $Y$ are the sets of x and y coordinates in the grid. For isotropic media FMM has often been used to calculate travel time fields, and works in a similar way to Dijkstra's algorithm to solve boundary problems. In this paper the FMM is used to solve the eikonal equation:
\begin{equation}
\label{eq:eikonal}
\vert \triangledown \tau(x,y) \vert = \frac{1}{v(x,y,  \triangledown \tau / | \triangledown \tau |)}
\end{equation}
where $\tau(x,y)$ is the shortest travel time from the source to the point $(x, y)$. In isotropic media the phase velocity is independent of the direction of travel, in which case $v(x,y, \triangledown \tau / | \triangledown \tau |)$ is the speed at $(x,y)$. For anisotropic media $v(x,y,  \triangledown \tau / | \triangledown \tau |)$ is the phase velocity at location $(x,y)$ in the direction which is normal to the wave front, which in turn is dependent on the source location. FMM works by giving each point one of three states: ``{known}'', ``close'' or ``far'' relative to the current wave front. When a grid point has a ``known'' state, the travel time is fixed. A ``close'' point, has a travel time assigned, however its value is not fixed and can be updated. A ``far'' point currently has no travel time. To propagate the wave front we take the point with lowest travel time from the set of points in the ``close'' state (times can be stored in a minimum heap for efficiency) and set it to a ``known'' point. We then update the travel times for all surrounding ``far'' and ``close'' points using a finite difference approximation. This will either involve updating the time for a ``close'' point or calculating a travel time for a ``far'' point and setting it to be a ``close'' point. This is repeated until all points are in the ``known'' state. This method differs from Dijkstra's algorithm when updating the times as an upwind finite difference approximation is used with multiple surrounding points with ``known'' or ``close'' states, while Dijkstra uses a single point and uses the travel time between two points. For locally anisotropic media  Anisotropic Multi-Stencil Fast Marching Method (AMSFMM) \cite{tant_effective_2020} was developed to calculate travel time fields. This method uses a series of stencils on a 25 point finite difference scheme, where the direction of the stencil arms determines the velocity used for the finite difference approximation. Since those directions are constant across the whole grid, the travel times in those directions are very accurate, but travel time estimates in other directions can have large errors, especially in strongly anisotropic media. This bias towards particular directions makes this method unsuitable for ray tracing: when the geometry is rotated to lie on a different grid, travel times and hence ray paths may vary greatly. The errors can be reduced by using more stencils, but this is computationally more expensive and requires points to be used further from the grid point where the travel time is being estimated. This may in turn require a finer grid to be used in order to reduce errors at interfaces between different materials. AMSFMM also requires $90^\circ$ rotational symmetry in the group velocities, while many materials have only $180^\circ$ rotational symmetry which limits the types of materials and applications to which it can be applied. For these reasons we seek a different finite difference scheme to update travel times in FMM. The method proposed herein, uses a finite difference scheme similar to that of Eaton 1993 \cite{eaton_finite_1993}, where an estimate of the wave front is calculated by linear interpolation to obtain points in space with equal travel time. Connecting such points produces an estimate of the wave front locally, which can then be used to estimate the travel time at new points on the grid. Since the estimated wave front can be oriented in any direction, we employ the full anisotropic velocity curve rather than a limited set of fixed directions as in AMSFMM. We also use a square grid instead of the hexagonal one used in Eaton 1993 \cite{eaton_finite_1993}. Use of the hexagonal grid means that the maximum angle between the estimated wavefront normal and the preferential wavefront direction is $30^\circ$, whereas in our framework this value is reduced to $22.5^\circ$ (see Section 4.1). Additionally, in applying a hexagonal grid, either the top/bottom or left/right boundaries of the domain do not align with the grid. In this case the maximum angle between the estimated wavefront normal and the preferential wavefront direction can reach $60^\circ$ , while with implementation of our triangular stencils this remains at $22.5^\circ$.

AMSFMM and ALI-FMM uses group and phase velocity curves which can be obtained by solving the Christoffel equation from a material's stiffness tensor and density. There are at most 21 (independent) stiffness tensor elements which are often written as a 6 x 6 symmetric matrix, but for many materials such as transversely isotropic media there are only a few independent elements.

\section{Group and Phase velocities}
\label{sect:velocities}
\subsection{Calculating velocity curves}
\begin{figure}
\centering
\includegraphics[scale=0.5]{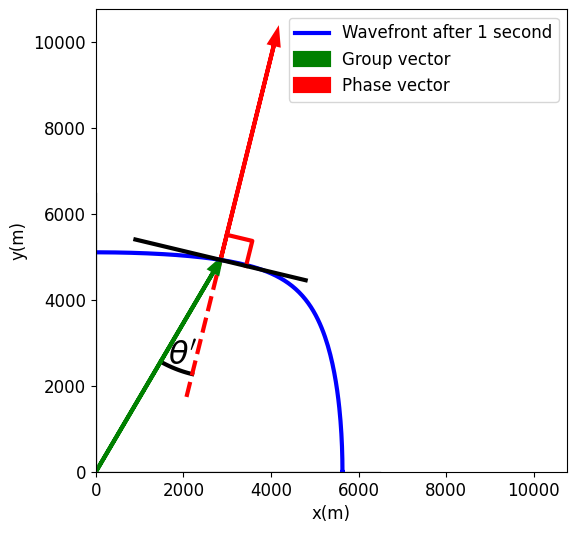}
\caption{Wave front after one second showing the group and phase vectors and the angle between them.}
\label{fig:wave_vectors}
\end{figure}

In anisotropic media, where the speed at which a wave travels is dependent on its direction of incidence, there are two different relevant and measurable velocities. The group velocity gives the speed at which a packet of energy comprising a band of neighbouring wave vectors travels in a certain direction, while the phase velocity gives the speed of wave fronts with individual wave vectors (Figure \ref{fig:wave_vectors}). For travel time tomography a material can be defined by group and phase velocity curves. These velocities can be derived from the Christoffel equation given a stiffness tensor and density, and so the group and phase velocity curves contain information about those parameters of the medium \cite{connolly_correction_2010}.

Since the velocity in anisotropic media is dependant on the angle of incidence, we want to determine the longitudinal or so-called P-wave (longitudinal is used for the remainder of this paper) group and phase velocities in a medium with constant stiffness tensor $c$ for different angles. Hooke's law \cite{JAEKEN2016445} requires that,
\begin{equation}
\sigma_{ij} = C_{ijql} \varepsilon_{ql}
\end{equation}
where $\sigma_{ij}$ is the stress tensor, $\varepsilon_{ql}$ is the elastic strain tensor and $C_{ijql}$ is the elastic stiffness tensor. Using the position vector $\mathbf{x}$ and displacement vector $\mathbf{u}=(u_i)$, Newton's second law can be rewritten in tensor notation as
\begin{equation}
\label{eq1}
\begin{split}
\rho \frac{\partial^2 u_i}{\partial t^2} & = C_{ijql} \frac{\partial \varepsilon_{ql}}{\partial x_j}\\
& = C_{ijql} \frac{\partial u_q}{\partial x_l \partial x_j}
\end{split}
\end{equation}
using the strain-displacement relationship and symmetry of the tensor notation we obtain Cauchy's law of motion, and thus we can look for a three dimensional plane wave solution of the form
\begin{equation}
u_i = Ap_i \textrm{exp} \left[ {\rm{i}} ( k_i x_i - \omega t) \right]
\end{equation}
where $A$ is a constant, $p_i$ are the components of the polarisation vector which describes the direction of the phase vector, $k_i = kn_i$ where $k$ is the wave number and $n_i$ is a component of the group vector direction $\mathbf{n}$.
The derivatives in equation (\ref{eq1}) are then given by,
\begin{equation}
\frac{\partial^2 u_i}{\partial t^2} = -\omega^2 u_i,
\end{equation}
\begin{equation}
\frac{\partial^2 u_q}{\partial x_j \partial x_l} = -k_jk_lu_q,
\end{equation}
Substituting these into equation (\ref{eq1}) gives
\begin{equation}
\begin{split}
\rho \omega^2 u_i & = C_{ijql}k_jk_lu_q\\
& = C_{ijql}k^2n_jn_lu_q
\end{split}
\end{equation}
which rearranges into the Christoffel equation
\begin{equation}
\label{Christoffel_eq}
\left[ \rho \omega^2 \delta_{iq} - C_{ijql} k^2 n_jn_l \right] \left[ u_k \right] = 0
\end{equation}
where $\delta_{ij}$ is the Kronecker delta function

\begin{equation}
\delta_{ij} = {\begin{cases}
1 , & \textrm{if } i=j \\
0 , & \textrm{otherwise}
\end{cases}}
\end{equation}
$\textbf{M}$ is the Christoffel matrix defined as $\textbf{M}_{ij} = C_{iqlj}k^2n_qn_l$. This can be solved as an eigenvalue problem to determine the phase velocities of all wave polarisation modes (quasi-longitudinal, quasi-shear horizontal and quasi-shear vertical) using equation (\ref{Christoffel_eq}). The wave front normal $n_j$ is given from the eigenvectors and the phase velocity used in equation (\ref{eq:eikonal}) is given by
\begin{equation}
v_i = \sqrt{\frac{\lambda_i}{\rho}}
\end{equation}
where $\lambda_i$ are the eigenvalues of $\textbf{M}$. Since longitudinal waves are the fastest, the largest eigenvalue corresponds to the longitudinal wave.
The group velocity is the speed in which a packet of energy goes in one direction and is given by
\begin{equation}
V_i = \frac{\partial \omega}{\partial k_i}
\end{equation}
We can multiply the Christoffel equation by $p_i$ to obtain
\begin{equation}
\begin{aligned}
&& \rho \omega^2 & =&& C_{ijql}k_qk_lp_jp_i\\
\Longrightarrow \; && \rho \frac{\partial (\omega^2)}{\partial k_m} & =&& C_{ijql}p_jp_i \frac{\partial (k_qk_l)}{\partial k_m}\\
\Longrightarrow \; && 2 \rho \omega \frac{\partial \omega}{\partial k_m} & =&& C_{ijql}p_jp_i \left( k_q\frac{\partial k_l}{\partial k_m} + k_l\frac{\partial (k_q)}{\partial k_m} \right)\\
&& & = && C_{ijql}p_jp_i (k_q \delta_{lm} + k_l \delta_{qm})\\
&& & = && C_{ijqm}p_jp_ik_q + C_{ijml}p_jp_ik_l
\end{aligned}
\end{equation}
which gives
\begin{equation}
V_m = \frac{\partial \omega}{\partial k_m} = \frac{C_{ijqm}p_jp_in_q}{\rho \omega}
\end{equation}
Alternatively since the group direction is the same as the wave front normal, we can determine the angle between the group and phase vectors (shown in Figure \ref{fig:wave_vectors}) and use that to determine the group velocity.
\begin{equation}
\label{eq:group_vel}
|V| = \frac{|v|}{\textrm{cos}(\theta')}
\end{equation}
where $\theta'$ is the angle between the group and phase vectors.

\subsection{Solving the Christoffel equation at run time}

In our forward model, materials can be defined by using a table of velocities at each integer angle; since changing the density of a material scales all velocities, the same table can be used to calculate the velocity for any density. However, if we want to use a variety of stiffness tensors, then we are limited to materials enumerated in the table of velocities. If we wish to use different tensors at each grid point this is potentially very computationally and memory intensive, especially if applied in 3D. This is because the Christoffel equation would have to be solved at every grid point, for every angle.\par
Instead we can solve the Christoffel equation at run time: this is computationally cheaper because the equation is solved for a single angle rather than generating the full velocity curve. Phase velocities are calculated using the method described above. Calculating group velocities is more difficult as it is calculated using the group angle rather than the phase angle. To achieve this we determine the phase angle from the group angle; the phase angle is then used to solve the Christoffel equation. We limit ourselves to materials in which wave energy that travels from a source to any other point on a given 2D plane will not go out plane (into the third dimension). We use orthotropic materials which have three orthogonal planes of symmetry; these include transversely isotropic, cubic and isotropic materials (for isotropic materials there is no need to solve the Christoffel equation as materials are defined by a single velocity). For transversely isotropic materials we can use the axis of symmetry to calculate velocities in 3D media. The stiffness matrix for these materials takes the form
\[ c = \begin{pmatrix}
c_{11} & c_{12} & c_{13} & 0 & 0 & 0\\
c_{12} & c_{22} & c_{23} & 0 & 0 & 0\\
c_{13} & c_{23} & c_{33} & 0 & 0 & 0\\
0 & 0 & 0 & c_{44} & 0 & 0\\
0 & 0 & 0 & 0 & c_{55} & 0\\
0 & 0 & 0 & 0 & 0 & c_{66}
\end{pmatrix} \]
where $c_{JL} = C_{ijkl}$ with $J=f(i,j)$ and $L=f(k,l)$ for
\[ f(i,j) = \begin{cases}
i, & \textrm{if }i=j\\
9-i-j, & \textrm{if }i \neq j
\end{cases} \]
In transversely isotropic media the stifness matrix is typically written with $c_{11} = c_{22}$, $c_{13} = c_{23}$ and $c_{55} = c_{66}$. Since this is isotropic in the XY plane, we are obtaining velocity curves from the YZ plane, therefore our phase vector is
\[ \begin{pmatrix}
0\\ \textrm{cos}(\theta) \\\textrm{sin}(\theta)
\end{pmatrix}\]
for phase angle $\theta$. We use this to determine the Christoffel matrix for a general phase angle which gives
\[ M = \begin{pmatrix}
\textrm{cos}^2(\theta)c_{66} + \textrm{sin}^2(\theta)c_{55} & 0 & 0\\
0 & \textrm{cos}^2(\theta)c_{22} + \textrm{sin}^2(\theta)c_{44} & \textrm{cos}(\theta)\textrm{sin}(\theta)(c_{23}+c_{44})\\
0 & \textrm{cos}(\theta)\textrm{sin}(\theta)(c_{23}+c_{44}) & \textrm{cos}^2(\theta)c_{44} + \textrm{sin}^2(\theta)c_{33}
\end{pmatrix} \]
One of the eigenvalues is $\textrm{cos}^2(\theta)c_{66} + \textrm{sin}^2(\theta)c_{55}$, however the eigenvector is $(1, 0, 0)'$ which corresponds to a shear wave which vibrates out of plane. The other two eigenvalues are calculated by solving a quadratic equation. The solutions are
\[ \frac{(\alpha + \gamma) \pm \sqrt{(\alpha - \gamma)^2 + 4\beta^2}}{2} \]
where $\alpha=\textrm{cos}^2(\theta)c_{22} + \textrm{sin}^2(\theta)c_{44}$, $\beta = \textrm{cos}(\theta)\textrm{sin}(\theta)(c_{23}+c_{44})$ and $\gamma = \textrm{cos}^2(\theta)c_{44} + \textrm{sin}^2(\theta)c_{33}$. Taking the largest root $\left(\left((\alpha + \gamma) + \sqrt{(\alpha - \gamma)^2+4\beta^2}\right)/2\right)$ gives the eigenvector for the longitudinal velocity since it is the fastest. We know that the eigenvector for the group velocity is
\[ X = \begin{pmatrix}
0\\ \textrm{cos}(\phi) \\\textrm{sin}(\phi)
\end{pmatrix}\]
for group angle $\phi$. The properties of eigenvectors are used to determine the phase angle for a fixed group angle by solving
\[ (M-\lambda I_3)X = 0\]
for some eigenvalue $\lambda \in \mathbb{R}$. There are two unknowns, $\theta$ and $\lambda$. Expanding the equation gives
\[ \begin{pmatrix}
\textrm{cos}^2(\theta)c_{66} + \textrm{sin}^2(\theta)c_{55} - \lambda & 0 & 0\\
0 & \textrm{cos}^2(\theta)c_{22} + \textrm{sin}^2(\theta)c_{44} - \lambda & \textrm{cos}(\theta)\textrm{sin}(\theta)(c_{23}+c_{44})\\
0 & \textrm{cos}(\theta)\textrm{sin}(\theta)(c_{23}+c_{44}) & \textrm{cos}^2(\theta)c_{44} + \textrm{sin}^2(\theta)c_{33} - \lambda
\end{pmatrix} \begin{pmatrix}
0\\ \textrm{cos}(\phi) \\\textrm{sin}(\phi)
\end{pmatrix} = \begin{pmatrix}
0\\ 0\\ 0
\end{pmatrix}\]
and hence the equations
\begin{equation}
0=0
\end{equation}
\begin{equation}
\label{eq:eigen_eq2}
(\textrm{cos}^2(\theta)c_{22} + \textrm{sin}^2(\theta)c_{44} - \lambda) \textrm{cos}(\phi) + \textrm{cos}(\theta)\textrm{sin}(\theta)(c_{23}+c_{44}) \textrm{sin}(\phi) = 0
\end{equation}
\begin{equation}
\label{eq:eigen_eq3}
\textrm{cos}(\theta)\textrm{sin}(\theta)(c_{23}+c_{44}) \textrm{cos}(\phi) + 
(\textrm{cos}^2(\theta)c_{44} + \textrm{sin}^2(\theta)c_{33} - \lambda ) \textrm{sin}(\phi) = 0
\end{equation}
By using the double angle formula and rearranging equations \ref{eq:eigen_eq2} and \ref{eq:eigen_eq3} we obtain
\begin{equation}
\label{eq:lambda_1}
2 \lambda = \textrm{cos}(2\theta) (c_{22} - c_{44}) + \textrm{sin}(2\theta)(c_{23}+c_{44}) \frac{\textrm{sin}(\phi)}{\textrm{cos}(\phi)} + (c_{22}+c_{44})
\end{equation}
\begin{equation}
\label{eq:lambda_2} 2\lambda = \textrm{cos}(2\theta) (c_{44}-c_{33}) + \textrm{sin}(2\theta)(c_{23}+c_{44}) \frac{\textrm{cos}(\phi)}{\textrm{sin}(\phi)} + (c_{44} + c_{33})
\end{equation}
We divide equations \ref{eq:lambda_1} and \ref{eq:lambda_2} by $\textrm{cos}(\phi)$ and $\textrm{sin}(\phi)$ which can not be done for angles which are multiples of $90^\circ$; however since both velocity curves are smooth functions and are symmetrical about these angles (materials are orthotropic), the group and phase angles must be equal in these directions. Equations \ref{eq:lambda_1} and \ref{eq:lambda_2} are used to eliminate $\lambda$ which gives
\[ \textrm{cos}(2\theta) (c_{22} + c_{33} - 2 c_{44}) + \textrm{sin}(2\theta)(c_{23}+c_{44}) \left( \frac{\textrm{sin}(\phi)}{\textrm{cos}(\phi)} -  \frac{\textrm{cos}(\phi)}{\textrm{sin}(\phi)} \right) + (c_{22} - c_{33}) = 0\]
Now let
\begin{equation}
\delta = c_{22} + c_{33} - 2 c_{44}
\end{equation}
\begin{equation}
\eta = (c_{23} + c_{44}) \left( \textrm{tan}( \phi ) - \frac{1}{\textrm{tan}(\phi )}\right)
\end{equation}
\begin{equation}
\kappa = c_{22} - c_{33}
\end{equation}
Since the equation is in the form $\delta$ cos$(2 \theta)$ + $\eta$ sin$(2 \theta)$ + $\kappa$, where $\delta,\eta,\kappa \in \mathbb{R}$, the substitution $y = \textrm{tan}(\frac{2 \theta}{2})$, where $\theta = \textrm{tan}^{-1}(y)$ is used to obtain sin$(2 \theta) = \frac{2y}{1+y^2}$ and cos$(2 \theta) = \frac{1-y^2}{1+y^2}$. This simplifies to a quadratic equation which we can easily solve
\begin{align*}
\delta \textrm{cos}(2\theta) + \eta \textrm{sin}(2\theta) + \kappa = & 0\\
\delta \frac{1-y^2}{1+y^2} + \eta \frac{2y}{1+y^2} + \kappa = & 0\\
y^2 (\kappa-\delta) + y (2\eta) + (\delta+\kappa) = & 0
\end{align*}
This gives $y = \frac{-\eta \pm \sqrt{\eta^2 + \delta^2 - \kappa^2}}{\kappa - \delta}$, and two solutions are obtained. However, one is for a shear wave mode and since we want the longitudinal wave mode the solution which gives the largest eigenvalue is used. For $\phi \in [0,180]$ the remainder is used since tan has multiple solutions when inverted (we seek angles between $0^\circ$ and $180^\circ$). For $\phi \in [0,180]$ we get the solution
\begin{equation}
\theta = \begin{cases} \begin{matrix}
\textrm{tan}^{-1} \left( \frac{-\eta - \sqrt{\eta^2 + \delta^2 - \kappa^2}}{\kappa - \delta} \right) \textrm{mod}\; 180 & \textrm{if } 0^\circ < \phi < 90^\circ \\
\textrm{tan}^{-1} \left( \frac{-\eta + \sqrt{\eta^2 + \delta^2 - \kappa^2}}{\kappa - \delta} \right) \textrm{mod}\; 180 & \textrm{if } 90^\circ < \phi < 180^\circ \\
\phi & \textrm{if } \phi \textrm{ mod } 90 = 0^\circ
\end{matrix}  \end{cases} 
\end{equation}

If $\phi \not \in [0,180]$ we can use the $180^\circ$ rotational symmetries to calculate the corresponding values, after which we obtain the phase velocity
\begin{equation}
|v| = \sqrt{\frac{\lambda}{\sigma}}
\end{equation}
with
\begin{equation}
\lambda = \begin{cases} \begin{matrix}
c_{22} & \textrm{if } \phi \; \textrm{mod}\; 180 = 0\\
c_{33} & \textrm{if } \phi \; \textrm{mod}\; 180 = 90\\
\frac{1}{2} \left[ \textrm{cos}(2\theta) (c_{22} - c_{44}) + \textrm{sin}(2\theta)(c_{23}+c_{44}) \textrm{tan}(\phi) + (c_{22}+c_{44})\right] & \textrm{otherwise} \end{matrix} \end{cases}
\end{equation}
The group velocity can also be obtained using equation (\ref{eq:group_vel}) where $\theta' = \theta - \phi$. For all of these equations, the only stiffness tensors required are $c_{22}, c_{23}, c_{33}$ and $c_{44}$ as the remaining stiffness tensors do not affect the velocities (when in plane). When these methods are implemented some of the numbers are very large and can cause overflow errors so we put stiffness tensors in MPa and scale velocities by 1000 to reduce the magnitude of numbers in the calculations. This scaling works for 64 bit numbers but does not for 32 bit numbers as the values are still too large. We also round angles when they are within $0.01^\circ$ of a multiple of 90 to bound these numbers which ensures the singularities of tangent function do not cause problems.

\section{Anisotropic Locally Interpolated Fast Marching Method (ALI-FMM)}
We now present a new algorithm based on the Fast Marching Method \cite{sethian_fast_1999} for calculating the travel time from a source to all points on a grid. ALI-FMM works by approximating the wave front using linear interpolation \cite{eaton_finite_1993} over a series of stencils which are used in an upwind finite difference approximation to propagate the wave front.
\subsection{Stencils}
We use a family of 32 finite difference stencils to approximate the wave front as locally planar, 16 of which form a square, and the other 16 form a triangle with a fourth grid point on the edge of the triangle. Three of the points in each stencil have travel time estimates and the other locates the point at which a travel time is to be estimated. We use the points with travel time estimates to calculate an estimate of the wave front. We begin by assigning labels to the stencil points. Tables \ref{tab:square_sten} and \ref{tab:triangle_sten} define points $A,B,C,D$ for square and triangular stencils, respectively, and diagrams showing the geometries of each stencil are given in the appendix (Figures \ref{fig:apendix_stencils1} and \ref{fig:apendix_stencils2}). In each case points $A,B$ and $C$ have travel time estimates and point $D$ is unknown, such that $\tau_A < \tau_B \leq \tau_C$, where $\tau_{*} = \tau(x_*,y_*, \triangledown \tau / | \triangledown \tau |)$. The later condition implies that stencils occur in pairs that use the same points for $A$ and $D$ but with positions of $B$ and $C$ swapped. When these conditions are not satisfied the stencil is rejected. 
\par

\begin{table}
\centering
\begin{tabular}{|l|l|l|l|l|}
\hline
Stencil & $A$ & $B$ & $C$ & $D$ \\ \hline
1-1 & $(i-1,j-1)$ & $(i,j-1)$ & $(i-1,j)$ & $(i,j)$ \\ \hline
1-2 & $(i-1,j-1)$ & $(i-1,j)$ & $(i,j-1)$ & $(i,j)$ \\ \hline
2-1 & $(i-1,j+1)$ & $(i,j+1)$ & $(i-1,j)$ & $(i,j)$ \\ \hline
2-2 & $(i-1,j+1)$ & $(i-1,j)$ & $(i,j+1)$ & $(i,j)$ \\ \hline
3-1 & $(i+1,j+1)$ & $(i+1,j)$ & $(i,j+1)$ & $(i,j)$ \\ \hline
3-2 & $(i+1,j+1)$ & $(i,j+1)$ & $(i+1,j)$ & $(i,j)$ \\ \hline
4-1 & $(i+1,j-1)$ & $(i,j-1)$ & $(i-1,j)$ & $(i,j)$ \\ \hline
4-2 & $(i+1,j-1)$ & $(i,j-1)$ & $(i+1,j)$ & $(i,j)$ \\ \hline
5-1 & $(i-2,j)$ & $(i-1,j-1)$ & $(i-1,j+1)$ & $(i,j)$ \\ \hline
5-2 & $(i-2,j)$ & $(i-1,j+1)$ & $(i-1,j-1)$ & $(i,j)$ \\ \hline
6-1 & $(i,j+2)$ & $(i+1,j+1)$ & $(i-1,j+1)$ & $(i,j)$ \\ \hline
6-2 & $(i,j+2)$ & $(i-1,j+1)$ & $(i+1,j+1)$ & $(i,j)$ \\ \hline
7-1 & $(i,j-2)$ & $(i+1,j-1)$ & $(i-1,j-1)$ & $(i,j)$ \\ \hline
7-2 & $(i,j-2)$ & $(i-1,j-1)$ & $(i+1,j-1)$ & $(i,j)$ \\ \hline
8-1 & $(i+2,j)$ & $(i+1,j+1)$ & $(i+1,j-1)$ & $(i,j)$ \\ \hline
8-2 & $(i+2,j)$ & $(i-1,j-1)$ & $(i+1,j+1)$ & $(i,j)$ \\ \hline
\end{tabular}
\caption{Labels of the points used in each of the square stencils for the finite difference approximation using grid coordinates $(i,j)$. Stencils are shown graphically in figure \ref{fig:apendix_stencils1}}
\label{tab:square_sten}
\end{table}

\begin{table}
\centering
\begin{tabular}{|l|l|l|l|l|}
\hline
Stencil & $A$ & $B$ & $C$ & $D$ \\ \hline
9-1 & $(i,j+2)$ & $(i,j+1)$ & $(i+1,j+1)$ & $(i,j)$ \\ \hline
9-2 & $(i,j+2)$ & $(i+1,j+1)$ & $(i,j+1)$ & $(i,j)$ \\ \hline
10-1 & $(i,j-2)$ & $(i,j-1)$ & $(i+1,j-1)$ & $(i,j)$ \\ \hline
10-2 & $(i,j-2)$ & $(i+1,j-1)$ & $(i,j-1)$ & $(i,j)$ \\ \hline
11-1 & $(i,j-2)$ & $(i,j-1)$ & $(i-1,j-1)$ & $(i,j)$ \\ \hline
11-2 & $(i,j-2)$ & $(i-1,j-1)$ & $(i,j-1)$ & $(i,j)$ \\ \hline
12-1 & $(i,j+2)$ & $(i,j+1)$ & $(i-1,j+1)$ & $(i,j)$ \\ \hline
12-2 & $(i,j+2)$ & $(i-1,j+1)$ & $(i,j+1)$ & $(i,j)$ \\ \hline
13-1 & $(i-2,j)$ & $(i-1,j)$ & $(i-1,j+1)$ & $(i,j)$ \\ \hline
13-2 & $(i-2,j)$ & $(i-1,j+1)$ & $(i-1,j)$ & $(i,j)$ \\ \hline
14-1 & $(i+2,j)$ & $(i+1,j)$ & $(i+1,j+1)$ & $(i,j)$ \\ \hline
14-2 & $(i+2,j)$ & $(i+1,j+1)$ & $(i+1,j)$ & $(i,j)$ \\ \hline
15-1 & $(i+2,j)$ & $(i+1,j)$ & $(i+1,j-1)$ & $(i,j)$ \\ \hline
15-2 & $(i+2,j)$ & $(i+1,j-1)$ & $(i+1,j)$ & $(i,j)$ \\ \hline
16-1 & $(i-2,j)$ & $(i-1,j)$ & $(i-1,j-1)$ & $(i,j)$ \\ \hline
16-2 & $(i-2,j)$ & $(i-1,j-1)$ & $(i-1,j)$ & $(i,j)$ \\ \hline
\end{tabular}
\caption{Labels of the points used in each of the triangular stencils for the finite difference approximation using grid coordinates $(i,j)$. Stencils are shown graphically in figure \ref{fig:apendix_stencils2}.}
\label{tab:triangle_sten}
\end{table}

\begin{figure}
\centering
\begin{subfigure}[b]{0.35\textwidth}
\includegraphics[scale=0.5]{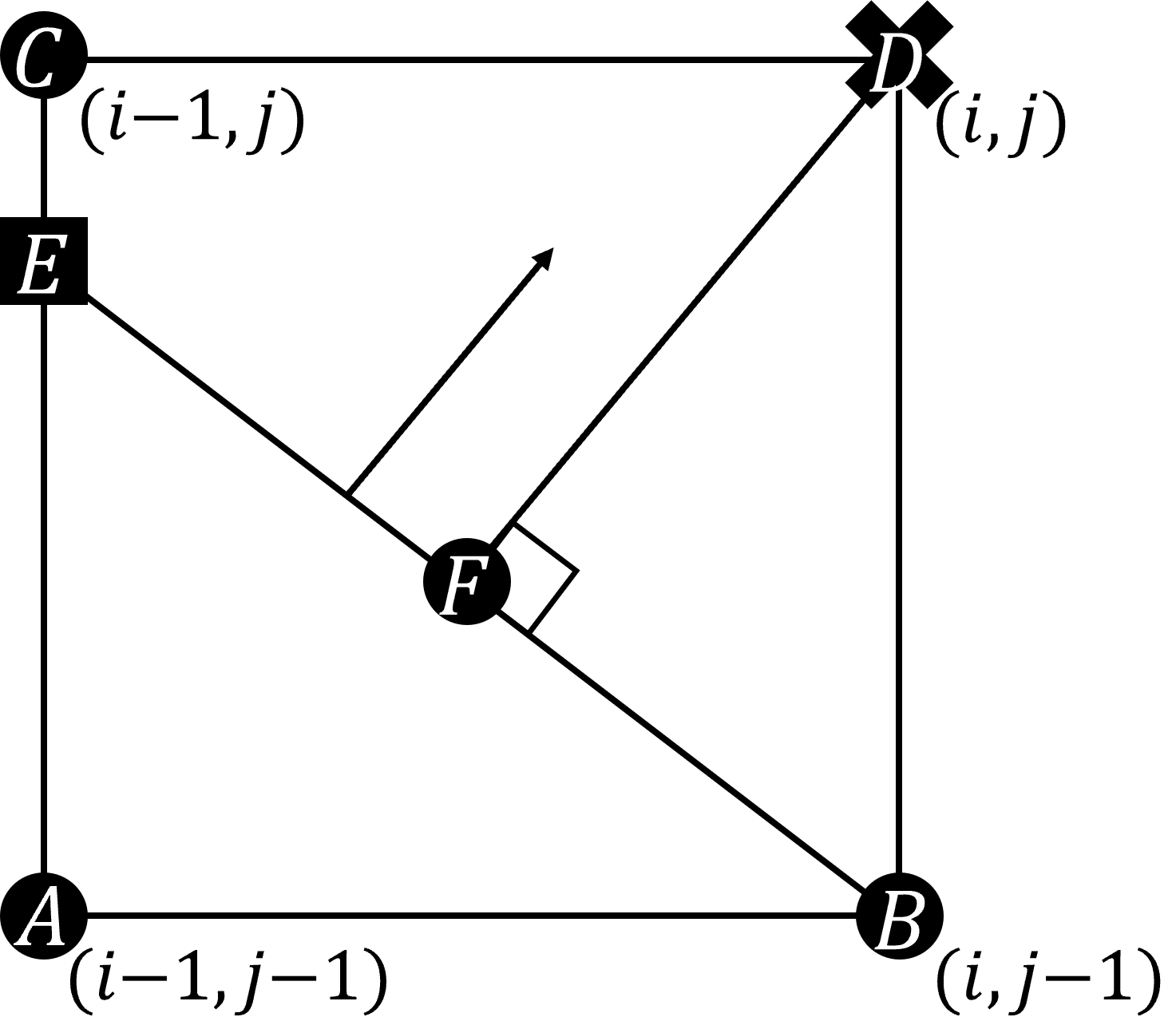}
\caption{stencil 1-1}
\end{subfigure}
\hspace{0.05\textwidth}
\begin{subfigure}[b]{0.55\textwidth}
\includegraphics[scale=0.5]{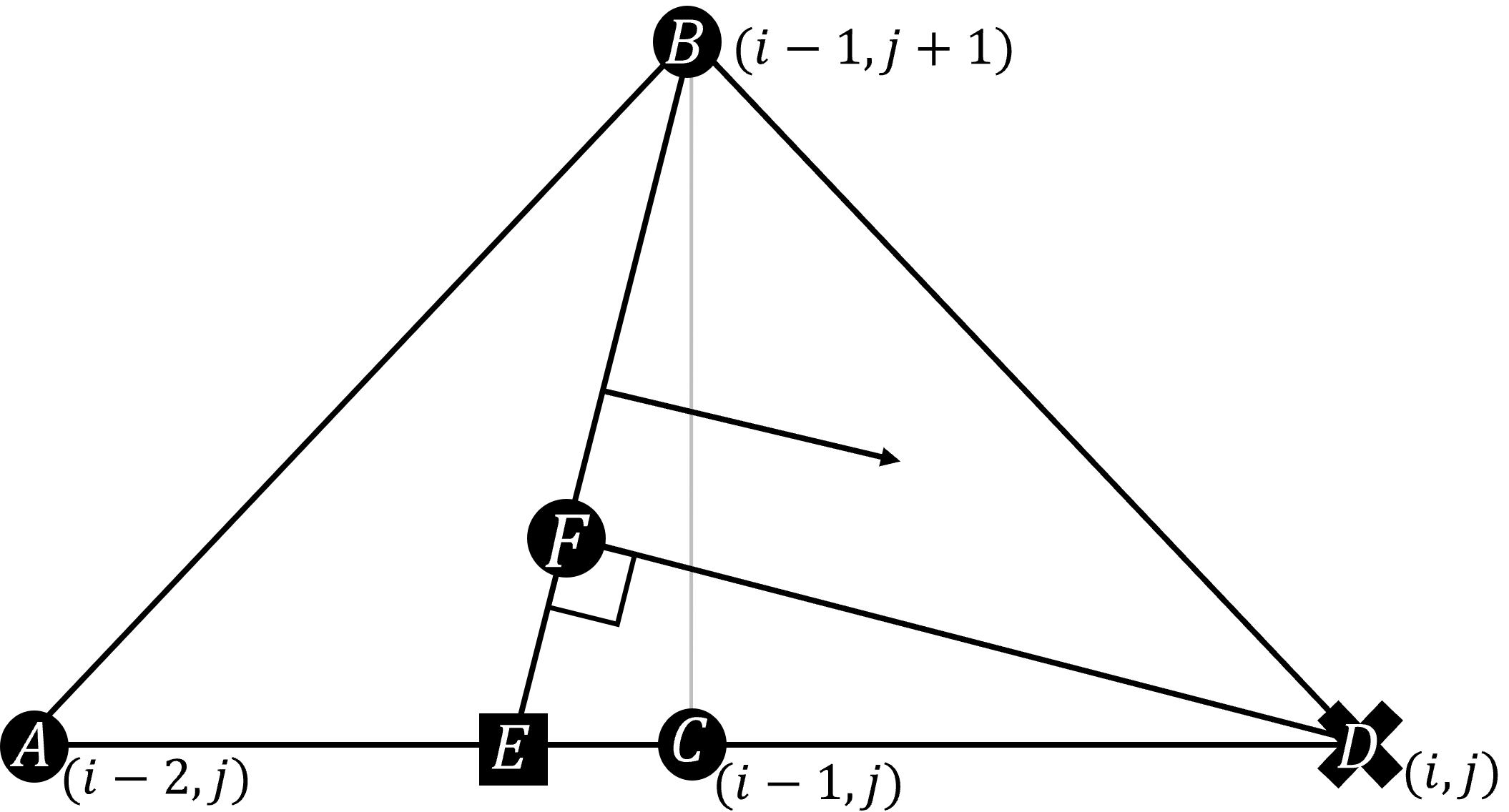}
\caption{stencil 13-2}
\end{subfigure}
\caption{Estimation of a locally linear wave front $EB$, a normal vector $FD$ and hence distance $|FD|$. The remaining stencils are included in Figures \ref{fig:apendix_stencils1} and \ref{fig:apendix_stencils2}.}
\label{Wave_front_est}
\end{figure}

We find the point $E$ on line $AC$ that has the same travel time as $B$ using linear interpolation of the travel times between $A$ and $C$:
\begin{equation}
\label{eq:wavefront_lin_int}
(x_E, y_E) = (1-a) (x_A, y_A) + a (x_C, y_C)
\end{equation}
for some $a \in [0, 1]$ subject to $\tau_B = (1-a)\tau_A + a \tau_C$, which is solved by $a = \frac{\tau_B-\tau_A}{\tau_C-\tau_A}$. We approximate the wavefront locally by the straight line joining $B$ and $E$ (Figure \ref{Wave_front_est}).

\subsection{Travel time estimate from wave front}
In order to calculate an estimate for the travel time at $D$ we solve the eikonal equation (Equation \ref{eq:eikonal}). For anisotropic media, $v(x,y,\triangledown \tau / | \triangledown \tau |)$ is the phase velocity at $(x,y)$ in the direction which is locally normal to the wave front. Since we have a locally linear wave front estimate we can use the orientation of its normal vector as an estimate to the direction of $\triangledown \tau(x,y)$, $| \triangledown \tau(x,y)|$ is obtained from the phase velocity curve using that direction. Combining these we get an estimate of $\triangledown \tau(x,y)$ which is used to determine a travel time estimate.

We start by finding the point $F$, which is the closest point to $D$ on line $EB$ (see Figure \ref{Wave_front_est}). The location of $F$ is obtained by finding the line perpendicular to $EB$ passing through $D$, and then finding the point of intersection between this line and the wavefront estimate. The distance and direction from $F$ to $D$ is then obtained. Line $EB$ can be written as
\begin{equation}
\label{line_f1}
E + \lambda_1 \overrightarrow{EB} = \begin{bmatrix}
x_E\\
y_E
\end{bmatrix}
+ \lambda_1
\begin{bmatrix}
x_B-x_E\\
y_B-y_E
\end{bmatrix}
\end{equation}
for $\lambda_1 \in \mathbb{R}$.
The direction of line $FD$ can be found by rotating $\overrightarrow{EB}$ by $90^\circ$ to obtain $\begin{bmatrix}
-(y_B-y_E)\\
x_B-x_E
\end{bmatrix}$.\\
The distance from $F$ to $D$ is
\begin{equation}
\| \overrightarrow{FD} \| = \frac{\left| (y_E-y_B)(x_D-x_E) + (x_B-x_E)(y_D-y_E)\right| }{\sqrt{(x_B-x_E)^2 + (y_B-y_E)^2}}.
\end{equation}
The normal to the wave front is the phase angle/vector. We therefore use the phase velocity and distance to determine the difference in travel times ($\tau_D-\tau_F$) between $F$ and $D$. Since line $EB$ is our approximation of the wave front we assume $\tau_F = \tau_B$ so
\begin{equation}
\tau_D - \tau_F \approx \frac{\| \overrightarrow{FD} \|}{v(x_F, y_F, \theta ')}
\end{equation}
where $v(x_F, y_F, \theta ')$ is the phase velocity of the material at $F$ at angle $\theta'$. For this we use the material orientation at $D$ and the direction of the normal to the wave front to calculate the effective angle $\theta' = (\theta_D - \theta_n + 90) \textrm{mod} 180$, where $\theta_D$ is the material orientation at $D$, $\theta_n$ is the direction of line $EB$ ($\theta_n \pm 90$ is the direction of the normal vector, perpendicular to the wave front). Since we assume the velocity has $180^\circ$ symmetry, we take the remainder. These velocities can either be obtained using linear interpolation with a table of velocities or by solving the Christoffel equation at run time:
\begin{equation}
\tau_D \approx \tau_B + \frac{\| \overrightarrow{FD} \|}{v(x_F, y_F, \theta ')}
\end{equation}
\subsection{Choosing which stencil to use}
Square stencils are often more accurate than the triangular stencils as there is less bias toward some directions due to the symmetry of the stencil. Triangular stencils are only used when there are insufficient points with travel time estimates to use a square stencil. This should only occur when there is high curvature on the wave front. Triangular stencils are also used when the grid point is at the edge of the grid (some of stencils 5 to 8 can then not be used as they require a point which is not on the grid). 
For these stencils we want the group vector from the estimated wave front to $D$ to intersect line $EB$ as this is the region where we have performed a wave front estimate. This is more likely to occur the closer point $F$ is to the middle of line $EB$. For square stencils, we want to use the stencil which has a wave front which is closest to line $CB$. This will also have lower error in the linear interpolation, and brings $F$ closer to $D$ making the estimation of $\tau_D - \tau_F$ smaller and subsequently the error. Since we get closer to line $CB$ the closer the travel times at $C$ and $B$ are, the stencil with the smallest difference in travel time between $C$ and $B$ is used. For triangular stencils we want the perpendicular bisector of points $B$ and $E$ to pass close to point $D$, so that $F$ is close to the middle of line $EB$. This can only happen for the second stencil in each pair and points $EDB$ form an isosceles triangle. This happens when the length of line $ED$ is $\sqrt{2}$ (using grid indexing), the length of line $AE$ is $2 - \sqrt{2}$ and the line $EC$ is $\sqrt{2}-1$. Therefore $a$ in equation (\ref{eq:wavefront_lin_int}) would equal $2-\sqrt{2}$. We therefore find the stencil where $\tau_B$ is closest to $(\sqrt{2} -1)\tau_A + (2 - \sqrt{2})\tau_C$ for the second stencil in a pair (same calculation for the first stencil in each pair except that the positions of $B$ and $C$ are swapped).\par

\begin{figure}
\centering
\includegraphics[scale=0.4]{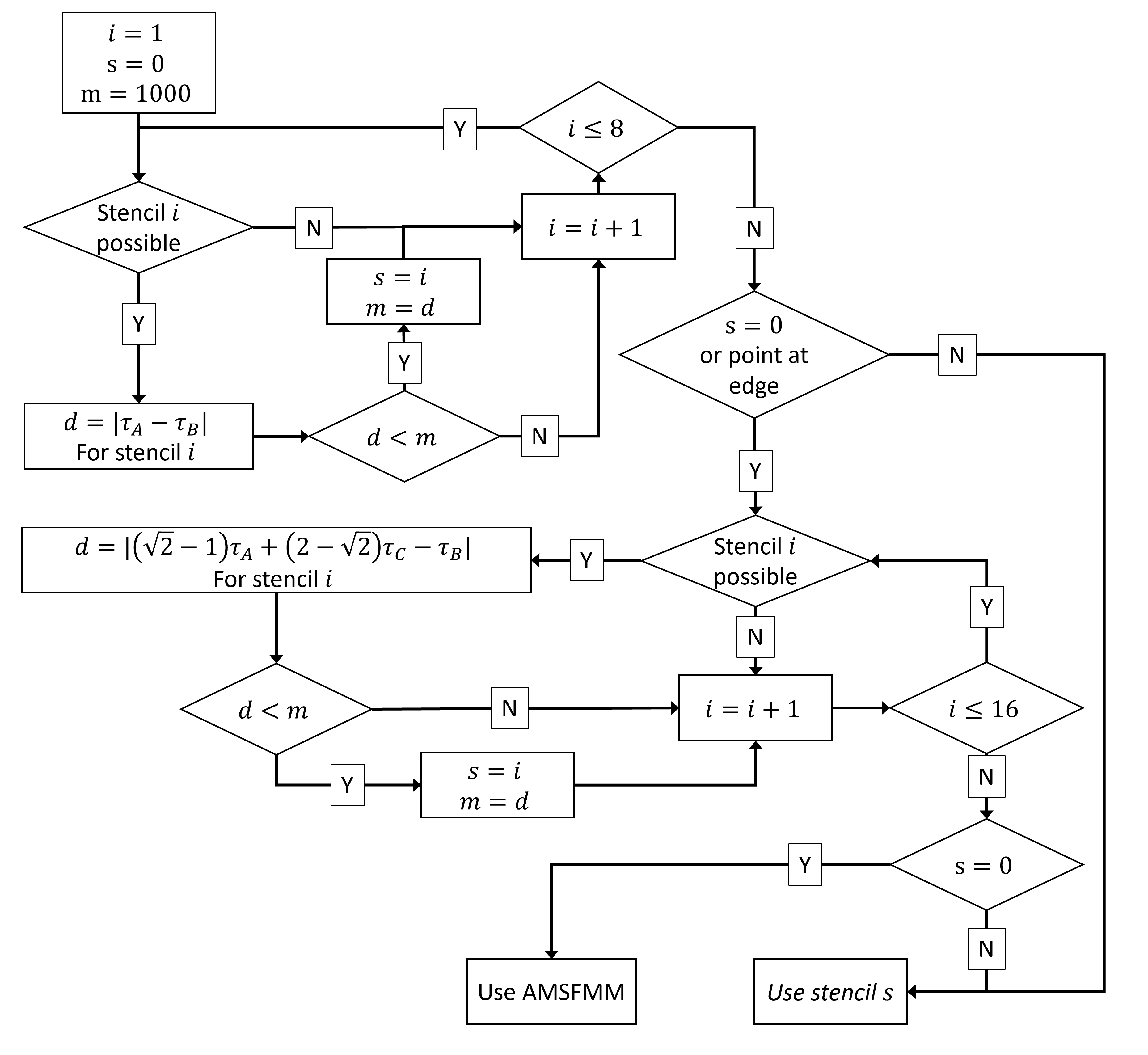}
\caption{Flowchart for choosing which stencil to use. Points $A,B,C,D$ are the points in the second stencil of each pair in the appendix (calculation is the same for the first stencil in each pair except position of $B$ and $C$ are swapped).}
\label{fig:choose_stencil}
\end{figure}

A stencil is only used when $\tau_D > \tau_A$, so no issues arise when the wave front is travelling in the opposite direction. There are two sizes of the square stencils (stencils 1 to 4 and 5 to 8), and where both are feasible the smaller stencils are more likely to be chosen because the difference in travel time between $C$ and $B$ is more likely to be smaller. These stencils use points closer together (the small square has side lengths of 1 and the larger one has length $\sqrt{2}$), meaning that the small stencils are often more accurate. The same applies when comparing stencils 1 to 4 and 9 to 16 when the grid point is on the edge of the grid. In rare cases in which no stencils can be used, the finite difference approximation from AMSFMM \cite{tant_effective_2020} is used. A flowchart for choosing which stencil to use is given in Figure \ref{fig:choose_stencil}.

\subsection{Propagation around source}
\begin{figure}
\centering
\includegraphics[scale=0.8]{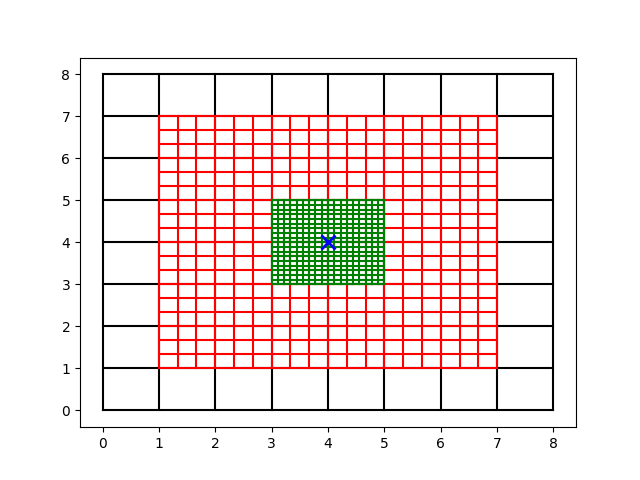}
\caption{Grids around source where intersection of lines are grid points with a source at the blue cross. Here we use a relative grid density of 9 around the source for a distance of 1(green region), before moving to a relative grid density of 3 at a distance of 3 (red region), and finally onto the original grid (black).}
\label{fig:source_grids}
\end{figure}

When wave propagation is initiated around a localised source, we commonly have too few grid points to represent the curvature of the wave fronts accurately. If the orientation of the medium's anisotropy is constant in space in the vicinity of the source we can use straight rays to calculate travel times out to a certain distance before changing to the finite difference method. Unfortunately we can not use this method for our applications in highly heterogeneous media. We therefore use a finer grid around the source (red and green grids in Figure \ref{fig:source_grids}) and assign orientations of the medium on the fine grid to be the orientation of the closest point on the original grid. The reason we do not use interpolation of the materials is because it would make our ray tracing method more computationally expensive since travel times of straight rays can not be calculated on the coarser grid. This would also cause issues around the source as it would not be homogeneous. We also use a very fine grid around the source (green grid in Fig \ref{fig:source_grids}) so that there is no change in the orientation map sufficiently close to the source (points with x and y coordinates from 3.5 to 4.5 in 
Figure \ref{fig:source_grids}). This allows us to use straight rays to calculate travel times to points on the latter grid. These points can then be used to begin wave front propagation using the finite difference method. Since using the finer grid is much more computationally expensive than the original grid, once the wave front reaches the boundary of a defined square region centred around the source, we move to a coarser grid and repeat this process  until we are on the original grid (see Figure \ref{fig:source_grids}). To create finer grids we increase the number of grid points in each direction by an odd number. An integer value means that we need not use interpolation to determine travel times on the coarser grid, and an odd number means that when assigning material properties to grid points there is a unique closest point on the original grid. We have used a factor of 3 for each grid refinement starting with a relative grid density of 27, and changing to a coarser grid when we calculate a travel time at a point which is at a distance of more than 1, then 6, and finally 13 grid points from the source (on the original grid) either horizontally or vertically. By using a finer grid and/or propagating further from the source before changing grids we can obtain more accurate travel times, but with an increased computational cost. This is because the change in wave front angle between grid points is smaller, so the approximation of the flat wave fronts (see Figure \ref{Wave_front_est}) is more accurate. For some applications we may want to keep the finer grid around the source instead of moving all points onto the coarser grid if there is a large difference in velocities spatially. This is because the wave front might not have moved far enough from the source in some directions, which would give inaccurate travel time estimates on the coarser grid due to the high curvature. In FMM all points neighbouring a ``known'' point are either ``known'' or ``close''. This ensures that the wave front continues to propagate. When moving to a coarser grid, ``known'' and ``far'' on the finer grid can be next to each other on the coarser grid which does not allow the wave front to propagate. Points which are ``known'' but have a neighbouring point which is ``far'' are added to the minimum heap (``close'' points are stored in the minimum heap, and when they are removed the surrounding ``close'' and ``far'' points are updated). This allows the neighbouring points to be changed to ``close'' points and obtain travel time estimates, while keeping the travel time at ``known'' points fixed. Source factorisation methods \cite{luo_fast_2012, desquilbet_single_2021, waheed2017} are an alternative approach for modelling around the source, however using grid refinement allows us to reduce the errors around the source by using finer grids or using them over a larger area (see Table \ref{tab:subgrid_errors}).

\subsection{Results}
We tested ALI-FMM using velocities obtained by solving the Christoffel equation for an austenitic steel, a cubic material with stiffness tensor elements $c_{11} = 203.6GPa$, $c_{12} = 133.5GPa$ and $c_{44} = 129.8GPa$ and density of $\sigma = 7850kg/m^3$ \cite{harvey_finite_elem}. The stiffness matrix for a cubic material takes the form
\[ c = \begin{pmatrix}
c_{11} & c_{12} & c_{12} & 0 & 0 & 0\\
c_{12} & c_{11} & c_{12} & 0 & 0 & 0\\
c_{12} & c_{12} & c_{11} & 0 & 0 & 0\\
0 & 0 & 0 & c_{44} & 0 & 0\\
0 & 0 & 0 & 0 & c_{44} & 0\\
0 & 0 & 0 & 0 & 0 & c_{44}
\end{pmatrix} \]
\begin{figure}
\centering
\includegraphics[width=.9\textwidth]{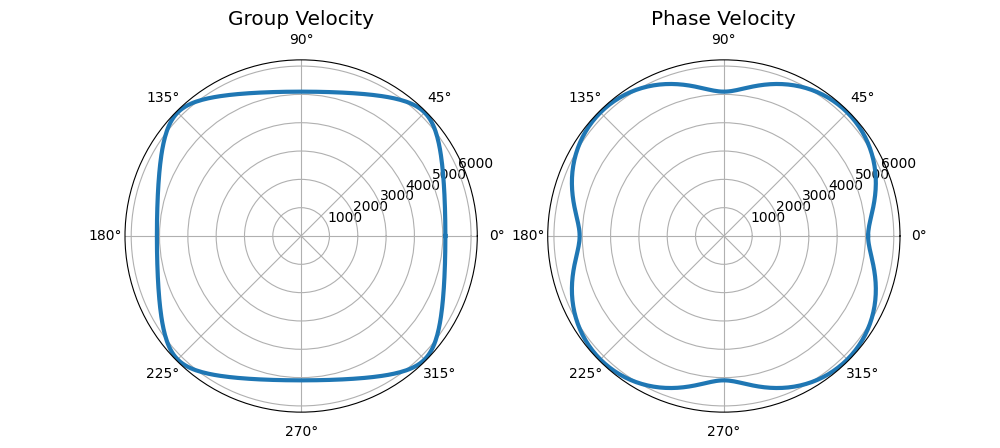}
\caption{Velocity curves for austenitic steel: (left) group velocity as a function of group angle, (right) Phase velocity as a function of phase angle.}
\label{fig:velocity_curves}
\end{figure}
and corresponding group and phase velocities are shown in Fig \ref{fig:velocity_curves}. In this example we know the true travel-times for homogeneous media (anisotropic orientations at every point are equal) which can be calculated using a straight ray from the source to any point on the grid, we can calculate the errors when using AMSFMM \cite{tant_effective_2020} and the new method herein.
\begin{figure}
 \begin{subfigure}{.5\textwidth}
  \centering
  \includegraphics[width=.9\textwidth]{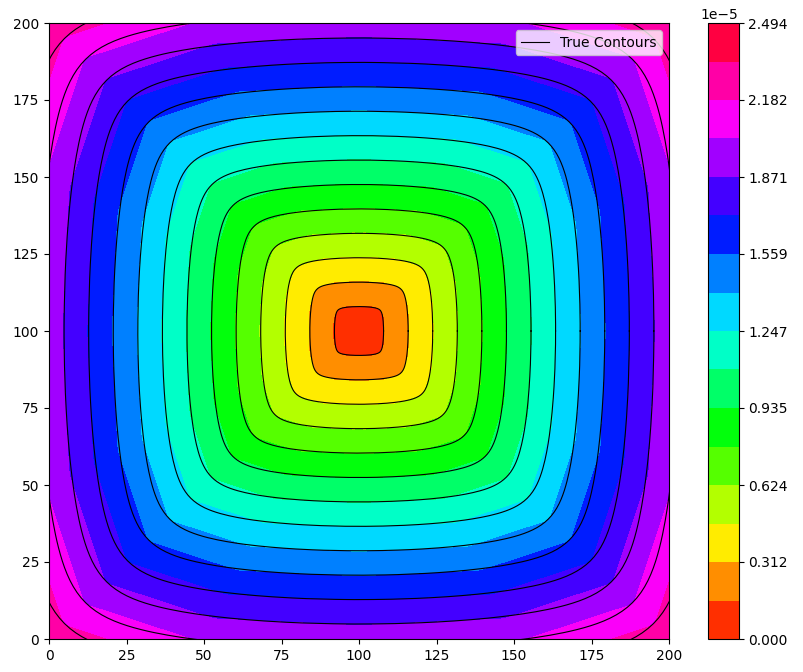}
  \caption{AMSFMM}
  \end{subfigure}
  \begin{subfigure}{.5\textwidth}
  \centering
  \includegraphics[width=.9\textwidth]{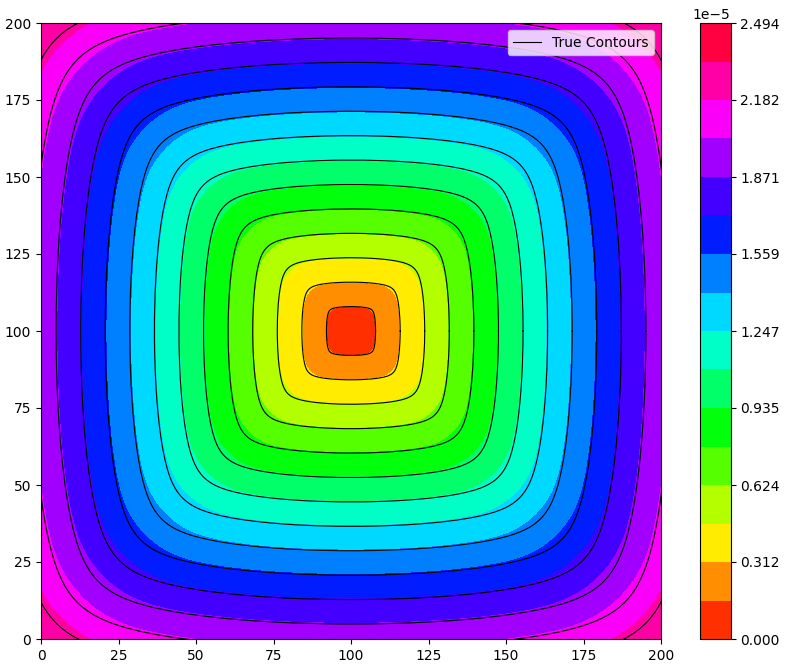}
  \caption{ALI-FMM}
  \end{subfigure}
  \caption{Contours of computed travel time field (colour) using AMSFMM (left) and ALI-FMM (right) with true contours (black) for a homogeneous medium with anisotropic orientation of $0^\circ$.}
  \label{fig:error_0deg}
\end{figure}
ALI-FMM commits the largest errors where the velocity has higher angular curvature. This is expected since approximating the wavefront locally by a straight line is less accurate for higher curvatures. In AMSFMM we have little  to no error in the directions of the stencil arms, however there is error between these directions. The errors are also very dependant upon the orientation of the medium relative to the grid (see Figures \ref{fig:error_0deg} and \ref{fig:error_36deg}).
\begin{figure}
 \begin{subfigure}{.5\textwidth}
  \centering
  \includegraphics[width=.9\textwidth]{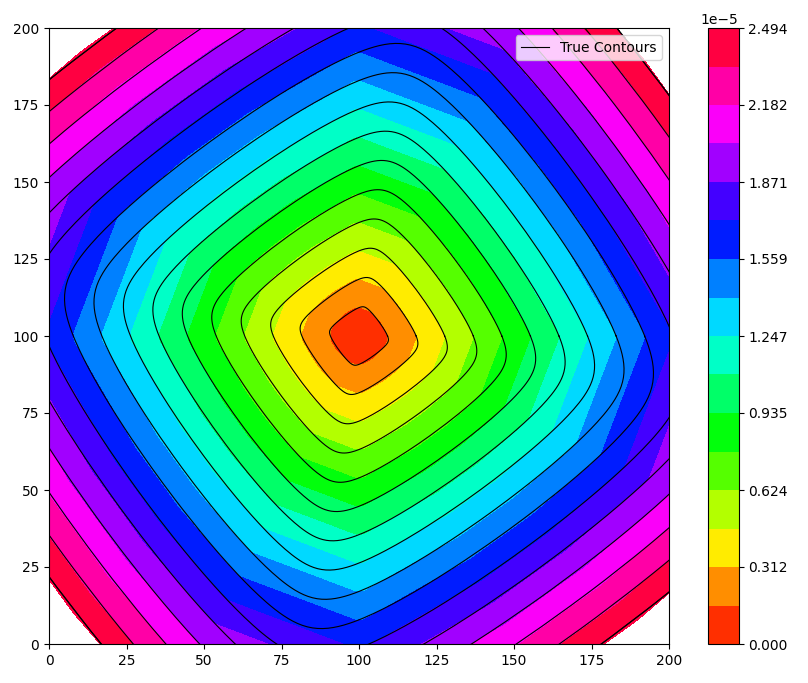}
  \caption{AMSFMM}
  \end{subfigure}
  \begin{subfigure}{.5\textwidth}
  \centering
  \includegraphics[width=.9\textwidth]{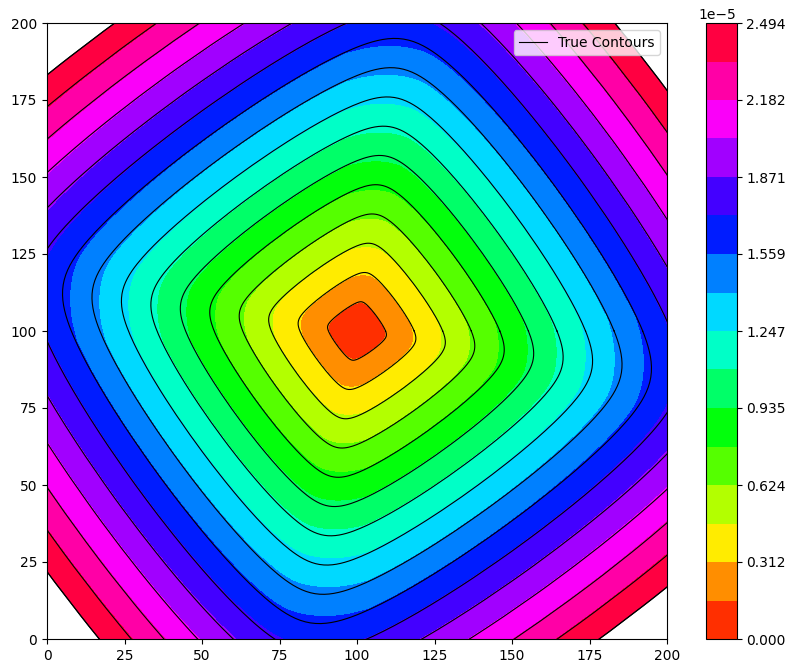}
  \caption{ALI-FMM}
  \end{subfigure}
  \caption{key simular to Figure \ref{fig:error_0deg} for a homogeneous medium with anisotropic orientation of $36^\circ$.}
  \label{fig:error_36deg}
\end{figure}
In Figure \ref{fig:error_36deg} the curvature of the wavefront is high between the directions of some of the stencil arms in AMSFMM which increases the error, in such orientations the ALI-FMM performs much better in these cases and has very similar error magnitudes to those in Figure \ref{fig:error_0deg}.
\begin{figure}
\centering
\includegraphics[width=.9\textwidth]{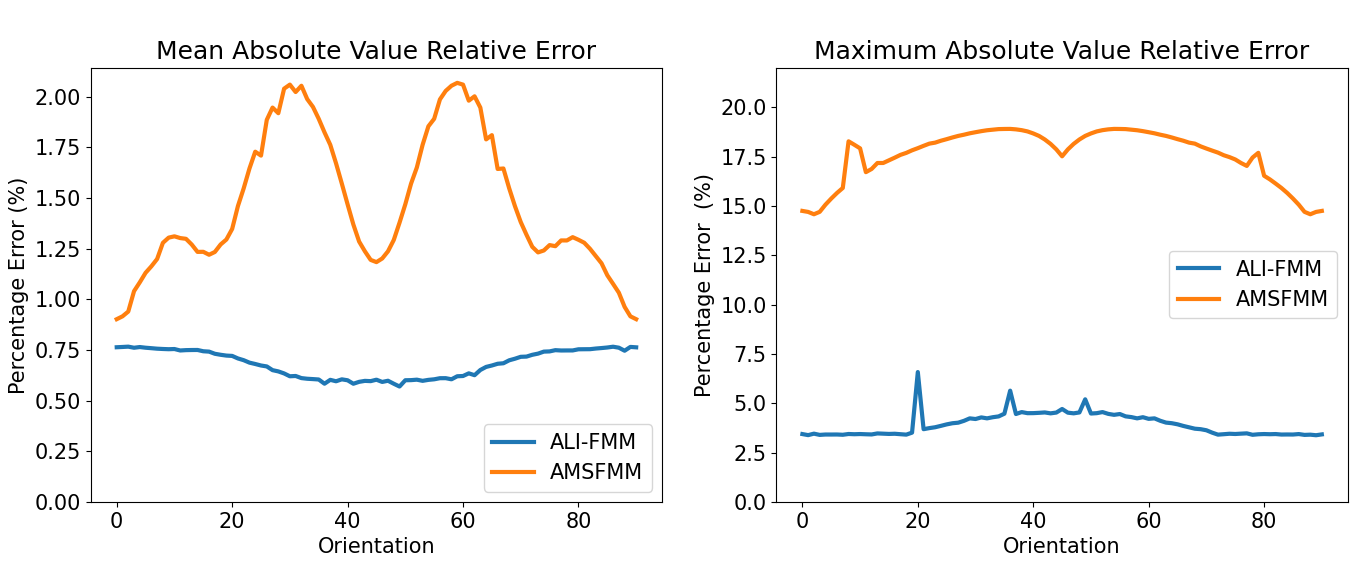}
\caption{(left) mean relative errors and (right) maximum relative error between computed travel time fields and the true solution for different anisotropic orientations in homogeneous anisotropic austenitic steel with group and phase velocities shown in figure \ref{fig:velocity_curves}.}
\label{fig:mean_max_error}
\end{figure}
Figure \ref{fig:mean_max_error} shows ALI-FMM has a lower and much more consistent average error, and the maximum error is approximately a factor of 4 smaller than AMSFMM for different orientations of this geometry.
\begin{figure}
\centering
\includegraphics[scale=0.6]{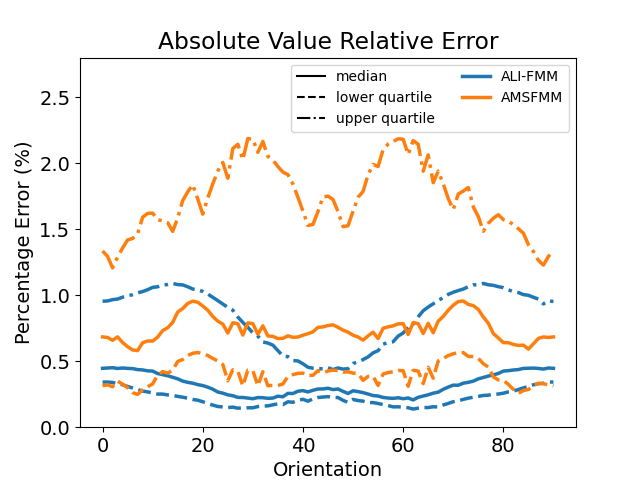}
\caption{Median, upper and lower quantiles for the relative error between the computed travel time fields and the true solution.}
\label{fig:error_quantiles}
\end{figure}
Figure \ref{fig:error_quantiles} shows that the median, upper and lower quartiles are lower in the new method except for a few angles of which we encounter similar errors in the lower quartile.

One way to improve the accuracy is to use a finer grid by adding additional grid points, so that the finite difference method is performed on a smaller area. When creating a finer grid we increase the number of grid points in each direction by a factor which we call the subgrid size which is an odd integer. We can assess the error for different sizes by using a homogeneous medium in which we can obtain the true travel times using straight rays. In order to have a fair comparison we only compare the travel times at points on the original grid. The original grid is 21 by 21 with the source at the center. The finer grids around the source remain the same as the previous results for a subgrid size of 1, however for the other sizes we use a relative grid density of 9 for a distance of 2 grid points, and a relative grid density of 3 for a distance of 5 grid points (distances are measured on the original grid). The grid used around the source is different to the original grid since we have already created a finer grid, hence there is a region around the source which has the same material properties as the source. An anisotropic orientation of $0^\circ$ was used as this had the lowest errors in the AMSFMM results.
\begin{figure}
\centering
\begin{subfigure}[b]{0.45\textwidth}
\includegraphics[scale=0.5]{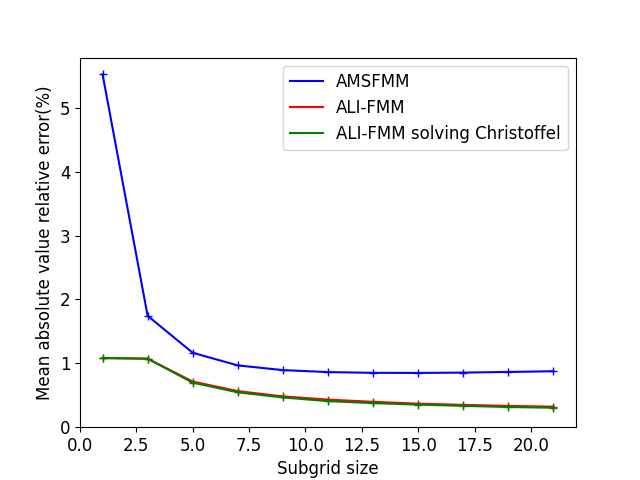}
\end{subfigure}
\hspace{0.05\textwidth}
\begin{subfigure}[b]{0.45\textwidth}
\includegraphics[scale=0.5]{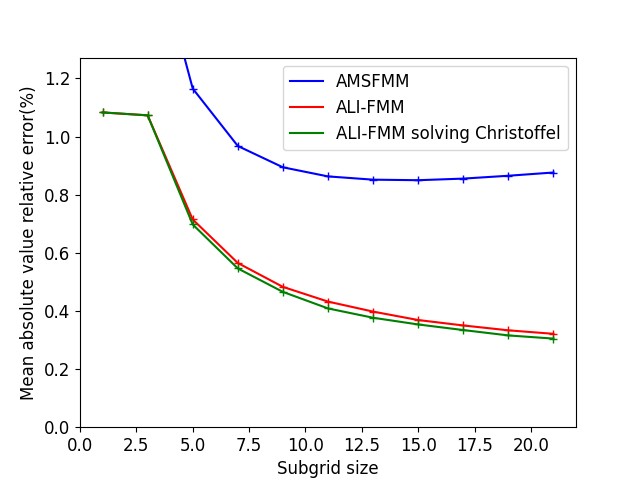}
\end{subfigure}
\caption{Mean absolute value percentage error for different methods (see legend) for different subgrid sizes with an anisotropic orientation of $0^\circ$.}
\end{figure}
\begin{center}
\begin{table}
\begin{tabular}{|l|l|l|l|l|l|l|l|l|l|l|l|}
\hline
Subgrid size & 1 & 3 & 5 & 7 & 9 & 11 & 13 & 15 & 17 & 19 & 21\\ \hline
AMSFMM & 5.527 & 1.744 & 1.166 & 0.968 & 0.894 & 0.863 & 0.852 & 0.850 & 0.855 & 0.865 & 0.876\\ \hline
ALI-FMM & 1.083 & 1.073 & 0.715 & 0.565 & 0.483 & 0.432 & 0.398 & 0.369 & 0.350 & 0.333 & 0.321\\ \hline
ALI-FMM & 1.083 & 1.073 & 0.698 & 0.546 & 0.466 & 0.409 & 0.377 & 0.353 & 0.334 & 0.316 & 0.305\\
solving Christoffel & & & & & & & & & & &\\
\hline
\end{tabular}
\caption{Absolute value relative percentage error on 21 by 21 grid using different subgrid sizes for the different methods.}
\label{tab:subgrid_errors}
\end{table}
\end{center}

Table \ref{tab:subgrid_errors} shows that the error in ALI-FMM is similar when using a subgrid size of 1 and 3. This is likely caused by the differences in the finer grids around the source. The errors for AMSFMM decrease quickly on the finer grid, however they start to increase after a subgrid size of 15. In ALI-FMM, these errors continue to decrease for finer grids and should indeed continue to decrease for all finer grids. The errors when we solve the Christoffel equation during run time are lower than those obtained using a table of velocities despite using the same methods. This is because the velocity calculations are more accurate as there is no linear interpolation between integer angles. Another way to increase accuracy is by using a finer grid over a larger area for the source which will have a similar affect to the subgrids, but only around the source.\\

All of the different codes use the same implementation of the Fast Marching Method for the binary tree and for storing the states (``far'', ``close'' and ``known'') of each point. These codes only differ in the finite difference method used and the finer grid around the source. This ensures that we can make a fair comparison of the compute times for different sizes of grids. A square grid with the source at the center has been used for benchmarking.
\begin{figure}
\centering
\begin{subfigure}[b]{0.45\textwidth}
\includegraphics[scale=0.5]{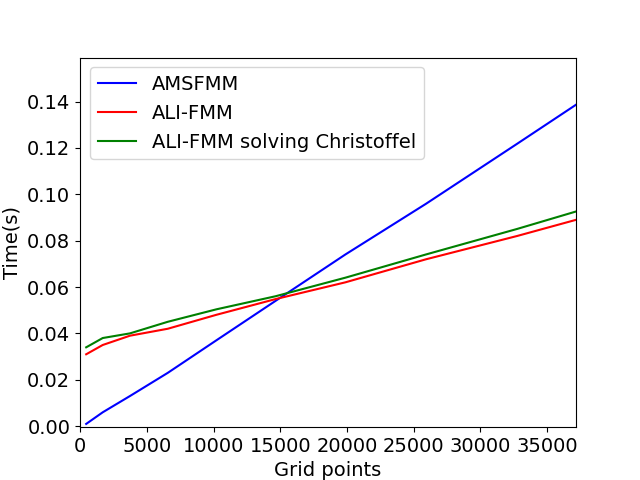}
\end{subfigure}
\hspace{0.05\textwidth}
\begin{subfigure}[b]{0.45\textwidth}
\includegraphics[scale=0.5]{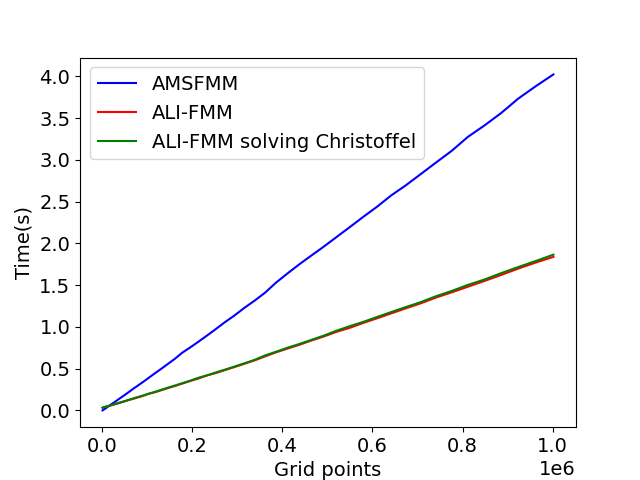}
\end{subfigure}
\caption{Times for computing travel time fields for different methods (see legend) for different sizes of grids (single threaded).}
\label{fig:compute_times}
\end{figure}
The methods have a complexity of $O(nm$ log$(nm)$) in Big O notation where $n$ and $m$ are the number of grid points in each direction. This complexity is because there are $nm$ points and we use a minimum heap structure to keep track of all points in the ``close'' state, which has logarithmic complexity for binary tree operations. However the operations in the heap are much faster than the finite difference methods making the complexity close to linear. Figure \ref{fig:compute_times} shows that AMSFMM is faster for small grid sizes, which is due to the finer grid around the source. Despite our finite difference method being slower than those used in AMSFMM, ALI-FMM is more efficient for larger grids. This is because when using AMSFMM, the stencil is chosen after the finite difference method is applied to all four stencils. ALI-FMM chooses the stencil before the finite difference method is applied to the stencil. As expected when we solve the Christoffel equation during run time, the time required is higher, but the added flexibility of using stiffness tensors at all grid points makes this method applicable to more complex problems.
\section{Ray Tracing in Anisotropic Media}

Finding the fastest ray path in isotropic media can be done by gradient descent through the travel time field from a receiver back to the source. This works because the ray paths are always locally perpendicular to the wave front in isotropic media. In anisotropic media the wave front is rarely perpendicular to the ray path so gradient descent will not produce accurate ray paths. Since the gradient gives the phase direction, the corresponding group direction is the direction of the ray path (see Figure \ref{fig:wave_vectors}), which can be obtained by solving the Christoffel equation (see Section \ref{sect:velocities}). This would requires us to know the relationship between group and phase directions which is unknown when using a table of velocities. In this section we show how travel time fields calculated above can nevertheless be used to find rays.

\subsection{Finding points along fastest ray paths}

\begin{figure}
\centering
\includegraphics[width=.9\textwidth]{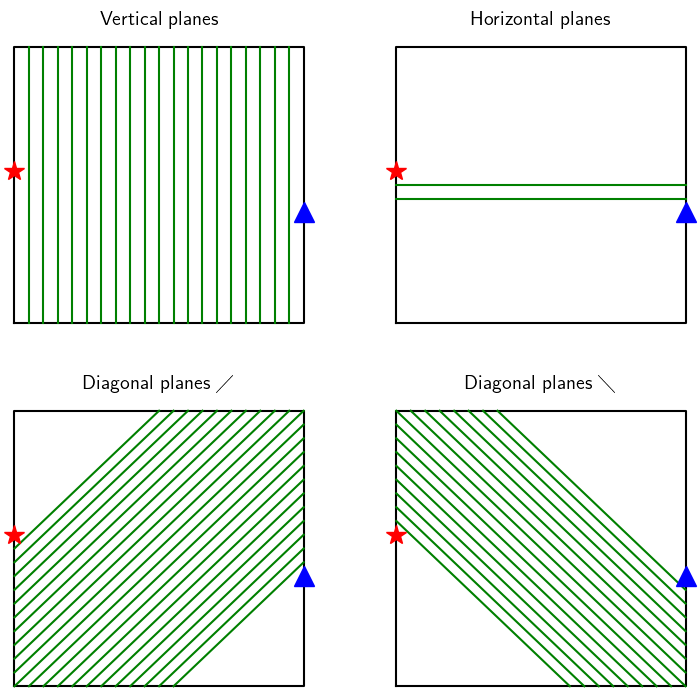}
\caption{Planes (green) used for ray path calculation between source (red star) and receiver (blue triangle)}
\label{Boundarys_plot}
\end{figure}

In order to find the fastest ray paths in anisotropic media, the grid is divided into sections by a series of planes. The ray path will be able to cross a plane once, and so the choice of these planes is important. The planes are chosen such that the points of intersection with the ray path have similar spacings. Since we use these points to parametrise the ray, we use equispaced horizontal, vertical or diagonal intersection planes passing through grid points, shown in Figure \ref{Boundarys_plot}.

We use four sets of planes because the choice of planes can affect the accuracy of the path. For example in Figure \ref{Boundarys_plot}, the horizontal planes are likely to perform poorly because we can only fit 2 planes between the source and receiver, providing a ray path approximated by only four points (source, receiver and two intersection points). In addition, the ray path can not travel above the source or below the receiver as all points must lie on a plane; in heterogeneous media we expect rays to deviate from such light constraints.

\begin{figure}
\centering
\includegraphics[scale=0.6]{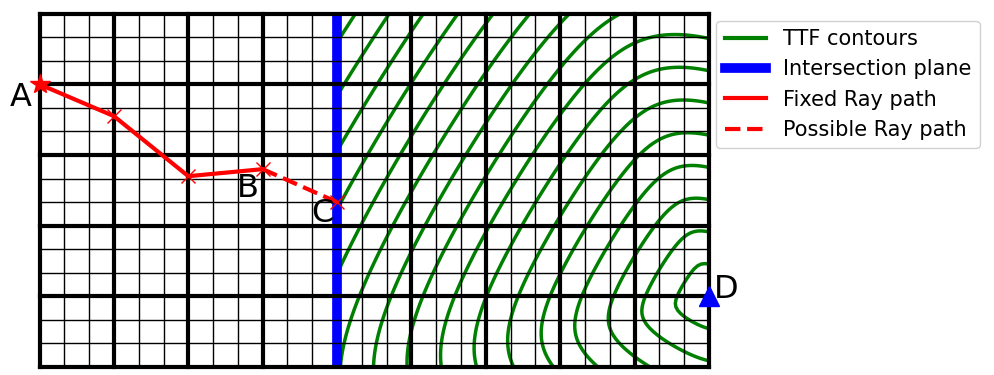}
\caption{The intersection between all thick black/blue lines are points from the original grid while all intersections of black and blue lines are points in the finer grid (In this case a 3x3 subgrid is used, but a finer grid is often used). The blue line is the plane where the next point on the ray path must lie. The red lines are sections of the ray path, where solid lines between $A$ and $B$ are fixed and the dashed line between $B$ and $C$ is a possible section of ray paths. The green lines are contours of the travel time field with source at $D$.}
\label{fig:ray_path_boundary}

\end{figure}

The spacing between planes is chosen so that the distance between intersection points is small enough to obtain sufficient detail about the ray path and large enough that the range of angles for lines $BC$ is sufficiently finely sampled on the boundary to get an accurate position for the next point. By putting the geometry onto a finer grid we can get a more accurate ray path which allows for a more detailed path. By using a finer grid we can use all possible planes on the original grid while having enough points for an accurate ray path (see Figure \ref{fig:ray_path_boundary}). The ray path is calculated using Fermat's principle, starting at the source and adding one intersection point at a time. Assuming that the first $n$ points on the ray path have been calculated, the next point is determined by finding the fastest path from the last calculated point on the ray path to the receiver. Let $A$ be the source, $B$ be the last calculated point on the ray path, $C$ be some 
grid point on the plane where the next point on the path may lie and $D$ is the receiver. Since the next point lies on a predetermined plane, we find the smallest travel time from $B$ to $D$ through $C$ shown in Figure \ref{fig:ray_path_boundary}. By finding the minimum total travel time of all points along the plane, we find the next point on the ray path. This is true because all other points on the plane result in a higher travel time and so by Fermat's principle the fastest ray path does not pass through these points.

The travel time from $B$ to $C$ is calculated assuming a straight ray, since our path will not have any points between them. The time can be obtained from the group velocity curve and the orientation of the closest point on the orientation map. For this step we can use the original grid as this is more efficient to calculate and since the geometry is identical, the travel time is unchanged. The fastest path from $C$ to $D$ is taken from the travel time field with $D$ as the source which can be obtained using ALI-FMM or any other method. This is the smallest travel time from $D$ to $C$, which is equal to the smallest travel time from $C$ to $D$, since the travel time is identical when source and receiver are swapped. The smallest travel time from $B$ to $D$ through $C$ under our constraints is the sum of these two travel times. Since the travel time field was only calculated at grid points, the minimum value on a plane is estimated by fitting a quadratic function to each local minimum and the two neighbouring points. We then find the minimum over these quadratic curves to estimate the global minimum. This allows the ray to travel to any point on the plane, instead of only to grid points giving a more accurate and smoother ray path. When we have multiple local minima there may be an alternate ray path, however this path will be slower or at best equally fast.

The choice of plane can make a difference to the accuracy of the ray path. However, we can calculate a ray path for each of the plane directions and integrate along the path to find the fastest time. The best results are often obtained when the planes are nearly perpendicular to the ray path. For more complex examples and to increase efficiency we can apply the same methods, but switch between plane directions based on the orientation of the last section of the ray. This allows us to obtain an accurate ray path when the ray intersects some planes more than once for all plane directions. This also reduces the computational cost as only one ray path is calculated. The plane direction is chosen by finding the plane for which the angle between the direction of the last section of the ray path and that plane is largest. For the initial point after the source, the direction from the source to the receiver is used. The position of the plane is calculated using the same spacing for the fixed plane directions, but using the distance between the plane and the last calculated point rounded to the nearest integer, to ensure that the plane passes through grid points. We stop when the distance between the last calculated point and the receiver is less than 1.6 grid points (on the original grid). This number is chosen arbitrarily, so that we do not pass the receiver and the spacing between the last two grid points is not too large. Since the angle between the direction of the ray and the plane is large, we can also limit the number of intersection points in the plane which are tested at each step. This does limit how much the ray can change direction, however with a sufficient number of points this should not effect the ray path.

\begin{figure}
\centering
\includegraphics[scale=0.8]{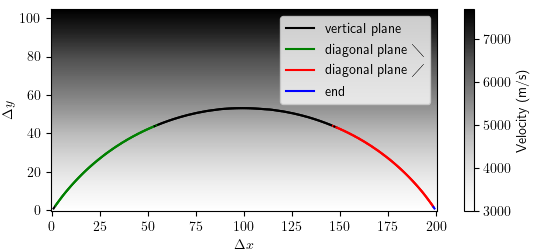}
\caption{Ray path in isotropic media with a velocity gradient showing plane directions used for ray tracing. Source is on the left and receiver is on the right.}
\label{fig:ray_directions}
\end{figure}
Figure \ref{fig:ray_directions} shows where each of the different plane directions are used in an example with a velocity gradient. The first point is calculated using a vertical plane (black); we then use three different plane directions as the direction of the ray changes, and finally add the last point in the ray when sufficiently close to the receiver (blue).

\subsection{Results}
All results from herein are obtained using grid spacings of 0.001 metres (1 mm) and using the same stiffness tensor as for the earlier results ($c_{11} = 203.6GPa$, $c_{12} = 133.5GPa$ and $c_{44} = 129.8GPa$ and density of $\sigma = 7850kg/m^3$). Ray tracing was first applied using travel time fields from AMSFMM \cite{tant_effective_2020}, but the error in travel-time fields is heavily biased towards low times in certain directions which cause issues in ray tracing. This often causes the ray paths to travel close to these directions and the four different planes give different ray paths due to this bias. These issues should be present for any ray tracing approach using these travel time fields unless the directional bias is compensated for.

\begin{figure}
\centering
\begin{subfigure}[b]{0.45\textwidth}
\includegraphics[scale=0.3]{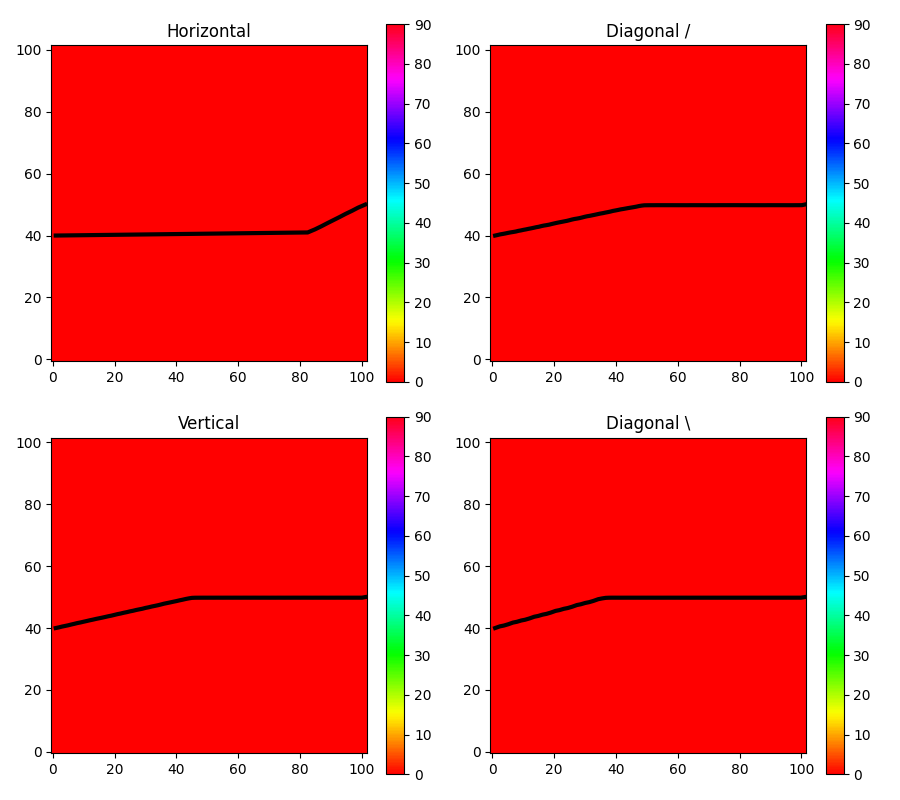}
\caption{AMSFMM}
\label{fig:homo_ray_using_AMSFMM}
\end{subfigure}
\hspace{0.03\textwidth}
\begin{subfigure}[b]{0.45\textwidth}
\includegraphics[scale=0.3]{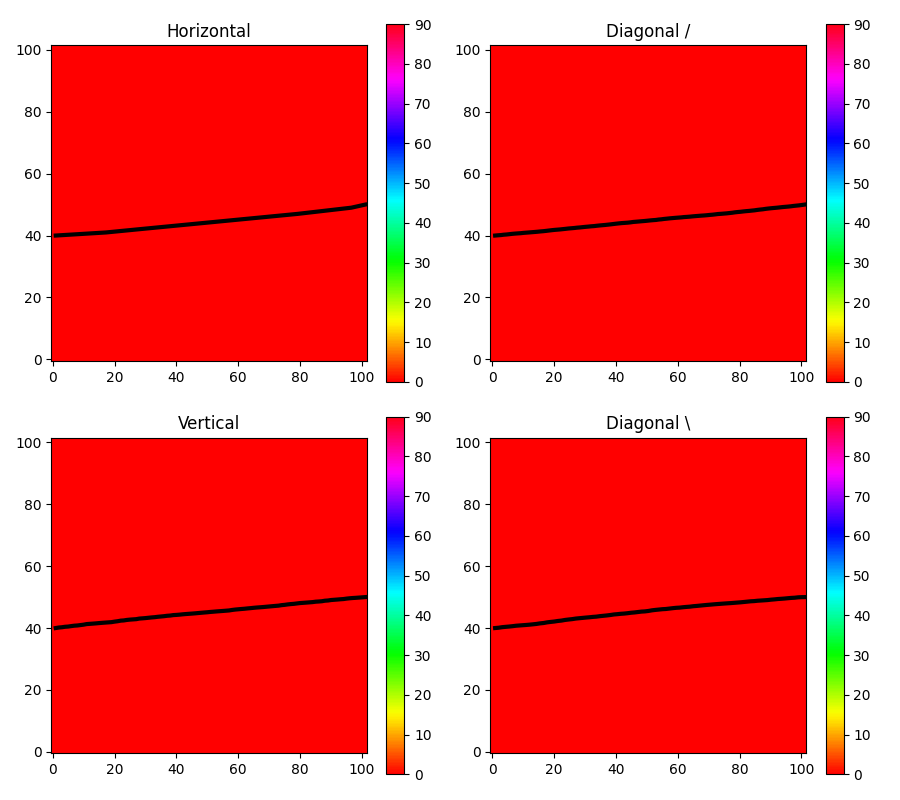}
\caption{ALI-FMM}
\label{fig:homo_ray_using_ALI_FMM}
\end{subfigure}
\caption{Ray paths calculated using travel time fields obtained from (a) AMSFMM, (b) ALI-FMM for the different planes in a homogeneous medium using a 9x9 subgrid. The background color is the orientation map used.}
\label{fig:homo_rays_using_AMSFMM_ALIFMM}
\end{figure}

In the homogeneous medium in Figure \ref{fig:homo_rays_using_AMSFMM_ALIFMM} the true solution is a straight ray from source to receiver (the anisotropic orientation is the same at all grid points). For AMSFMM the ray using the horizontal planes has the worst fit, which is to be expected as the angle between the direction of travel and the planes are much closer than for the other planes. This also means that there are fewer points in the ray path. Even though these ray paths are not straight, the travel times along these paths are lower than AMSFMM, except for the ray obtained using horizontal boundaries. Travel times of ray paths are generally lower than those obtained directly from AMSFMM travel time fields despite being calculated from AMSFMM travel time fields, with exceptions when the planes do not fit the path well or the true ray path is aligned along stencil directions (Table \ref{tab:ray_times}). This reduction in error is likely due to the ray tracing using the full velocity curve to calculate the travel time along line $BC$, which makes it more likely to travel in directions between stencil arms where the travel time field predicts the wavefront is slower than the true solution. This means that the ray path has less bias towards the directions of stencil arms than the travel time fields. The ray paths obtained using ALI-FMM are much straighter and all give similar paths for each of the plane directions as the travel times fields have less bias than AMSFMM.

\begin{figure}
\centering
\begin{subfigure}[b]{0.45\textwidth}
\includegraphics[scale=0.3]{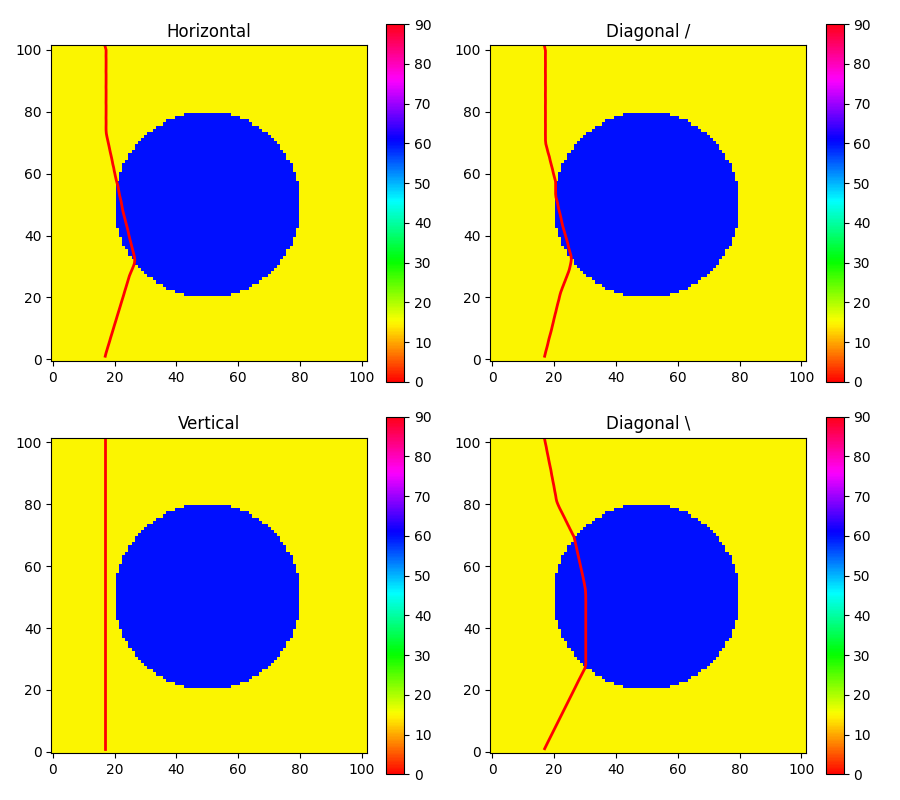}
\caption{AMSFMM}
\label{fig:ray_using_AMSFMM}
\end{subfigure}
\hspace{0.03\textwidth}
\begin{subfigure}[b]{0.45\textwidth}
\includegraphics[scale=0.3]{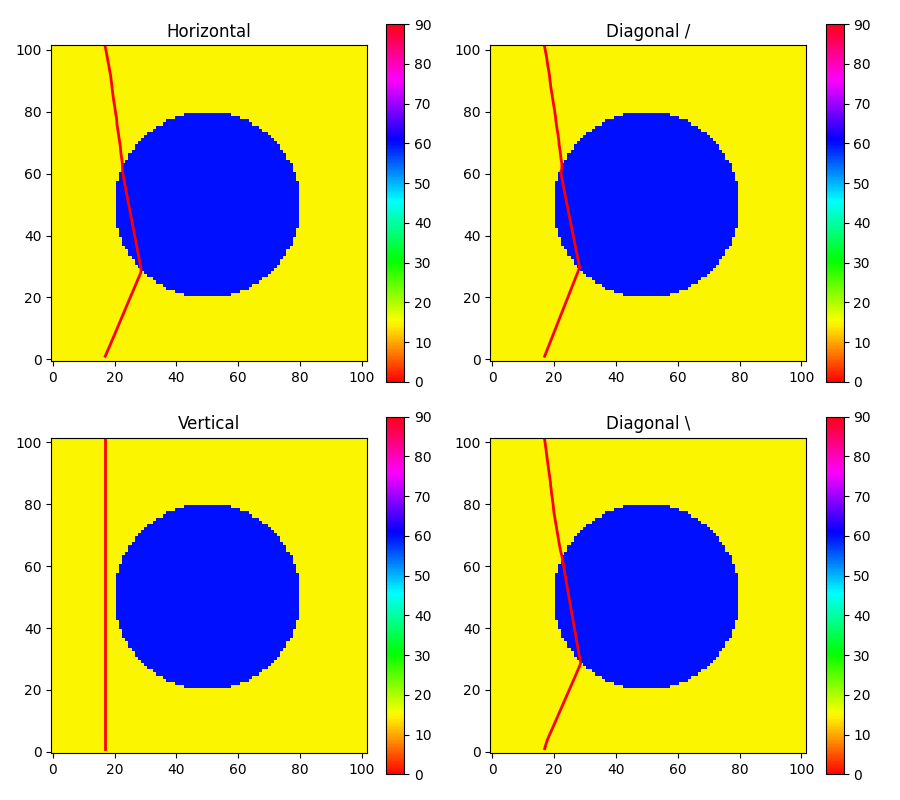}
\caption{ALI-FMM}
\label{fig:ray_using_ALI_FMM}
\end{subfigure}
\caption{Ray paths calculated using travel time fields obtained from (a) AMSFMM, (b) ALI-FMM for the different planes with a circular anomaly using a 9x9 subgrid. The background color is the orientation map used.}
\label{fig:rays_using_AMSFMM_ALIFMM}
\label{fig:circle_rays}
\end{figure}

\begin{figure}
\centering
\includegraphics[scale=0.7]{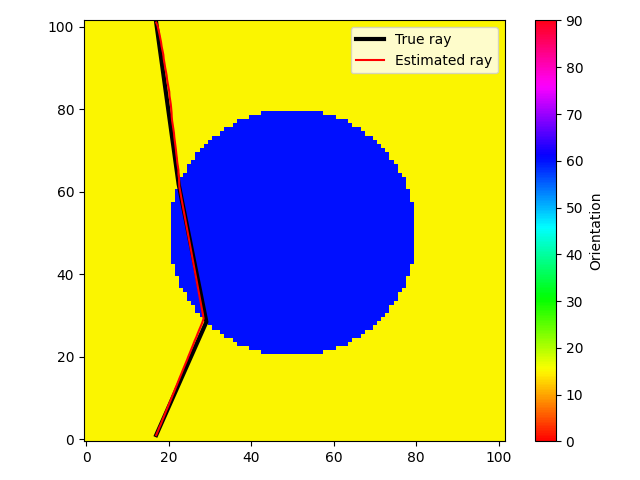}
\caption{Ray path calculated using travel time fields obtained from ALI-FMM and changing plane directions with a circular anomaly using a 9x9 subgrid. The background is the orientation map used.}
\label{fig:ray_changing_plane}
\end{figure}

The ray tracing method has also been applied with a circular anomaly using the same stiffness tensor we have used previously but with anisotropic orientations of $60^\circ$ inside and $15^\circ$ outside of the circle. In Figure \ref{fig:rays_using_AMSFMM_ALIFMM} the source and receiver have the same x coordinate so there are no vertical planes between them. This means that the ray paths obtained using vertical planes consists of only two points which are the source and receiver. For the other plane directions, the rays obtained using AMSFMM are biased towards certain directions and give different paths for the different plane directions due to the errors in the travel time fields similar to Figure \ref{fig:homo_ray_using_AMSFMM}, while the rays using ALI-FMM have far less bias and the different plane directions give similar paths and are much straighter. We then constructed a ray path by changing boundary orientation along the ray as described above (Figure \ref{fig:ray_changing_plane}). This method was only tested using ALI-FMM as ray tracing using AMSFMM is heavily biased and gives far less accurate ray paths due to the error in the travel time fields.
The corresponding ray path in Figure \ref{fig:ray_changing_plane} is similar to those obtained using a fixed plane direction, but since the plane direction may change along the ray, the problem that occurs when using a vertical boundary is unlikely to occur.
\begin{table}[h]
\begin{center}
\begin{tabular}{|l|c|c|}
\hline
Traveltime estimate & AMSFMM & ALI-FMM \\ \hline
Travel time field & 1.883465e-05 & 1.829758e-05 \\ \hline
Ray using horizontal planes & 1.841065e-05 & 1.828177e-05 \\ \hline
Ray using vertical planes & 1.912709e-05 & 1.912709e-05 \\ \hline
Ray using diagonal / planes & 1.847590e-05 & 1.831111e-05 \\ \hline
Ray using diagonal $\backslash$ planes & 1.858326e-05 & 1.826748e-05 \\ \hline
Ray using changing plane direction & & 1.827979e-05 \\ \hline \hline
'True' ray (circular geometry) & \multicolumn{2}{|c|}{1.828731e-05} \\ \hline
Optimal ray (discretised geometry) & \multicolumn{2}{|c|}{1.826223e-05} \\ \hline
\end{tabular}
\end{center}
\caption{Travel time estimates from travel time fields, and ray paths from Figures \ref{fig:rays_using_AMSFMM_ALIFMM} and \ref{fig:ray_changing_plane} (to 7 significant figures).}
\label{tab:ray_times}
\end{table}

Integrating travel times along the ray path will always give a time that is greater than or equal to the true travel time obtained on the discretized grid. The exact optimal path can be computed exactly. Assume that the disk where the material coefficients are modified has radius one, so that its boundary can be parameterized as (cos $\theta$,sin $\theta$), $\theta \in [0, 2 \pi]$. Then the optimal path is the concatenation of three straight segments, from $(x_0, y_0)$ the source point, to (cos $\theta_1$,sin $\theta_1$) a first point on the modified region boundary, to (cos $\theta_2$,sin $\theta_2$) a second point on the same boundary, to the target point $(x_1, y_1)$. The travel times along each of these segments can be computed using the material properties and the expressions of section 3. Then optimizing over $\theta_1, \theta_2 \in [0, 2 \pi]$ one can obtain the optimal path. This path (referred to as the 'True Path')  is plotted in Figure \ref{fig:ray_changing_plane} and produces a travel time of 1.828731e-05s. Note this 'true' travel time is in fact greater than that calculated using our proposed methods. This is because the discretisation of the underlying geometry (specifically the staircasing of the circle boundary) permits the ray to travel in preferential directions for longer. The optimal solution on this discretised geometry (obtained using a brute force approach) produces the shortest traveltime across all methods (1.826223e-05s to 7 significant figures). Since the rays obtained in Figure \ref{fig:rays_using_AMSFMM_ALIFMM} using horizontal planes gives a straight ray, this is by far the slowest ray path and shows that even though we use the same material throughout, the fastest path is not always a straight line from source to receiver since we have an anisotropic medium with varying anisotropic orientation. Table \ref{tab:ray_times} also shows that a large improvement in the travel time estimates can be obtained when using ray tracing with AMSFMM in comparison to times obtained directly from the travel time field. ALI-FMM also manages to obtain lower travel times (less error) for all estimates of the travel time compared to AMSFMM, except for the vertical planes which have already been discussed.

In anisotropic media there are only a few examples where the true solution is known other than for homogeneous media. One way to construct an example is to deform a homogeneous medium using a change of coordinates \cite{desquilbet_single_2021}. In isotropic media there are some examples where the true solution is known. One of these is when there is a constant velocity gradient in which case the solution is a section of a circle \cite{velocity_gradient_2001}. When the gradient is in the form $(g, 0)$, with source $(x_0,y_0)$ and receiver $(x_1,y_1)$, the circle has equation
\begin{equation}
\label{eq:circle_vel_grad}
\left( y' - \frac{v_0 \textrm{cot} \theta_0}{g} \right)^2 + \left( x' + \frac{v_0}{g} \right)^2 = \left( \frac{v_0}{g \textrm{sin} \theta_0} \right)^2
\end{equation}
where $x' = x - x_0$, $y' = y-y_0$ (source is at $(x',y') = $(0,0)), $v_0$ is the velocity at the source and $\theta_0$ is the initial ray direction relative to the velocity gradient. Unfortunately the value of $\theta_0$ is unknown, however the location of the receiver is known and can be used to find the location of the circle. Equation (\ref{eq:circle_vel_grad}) is used to solve for $\theta_0$:
\begin{align*}
\left( y' - \frac{v_0 \textrm{cot} \theta_0}{g} \right)^2 + \left( x' + \frac{v_0}{g} \right)^2 & = \left( \frac{v_0}{g \textrm{sin} \theta_0} \right)^2 \\
\left( gy' - v_0 \textrm{cot} \theta_0 \right)^2 + \left( gx' + v_0 \right)^2 & = v_0^2 
\left(1+\textrm{cot}^2 \theta_0 \right) \\
(gy')^2 + \left( gx' + v_0 \right)^2 -v_0^2 & = gy'v_0 
\textrm{cot} \theta_0 \\
\textrm{tan}^{-1}\left( \frac{gy'v_0}{(gy')^2+(gx'-v_0)^2-v_0^2} \right) & = \theta_0
\end{align*}
Therefore
\begin{equation}
\theta_0 = \textrm{tan}^{-1}\left( \frac{g(y_1-y_0)v_0}{(g(y_1-y_0)^2+(g(x_1-x_0)-v_0)^2-v_0^2} \right)
\end{equation}
From the equation of the circle, the center is at $(x',y')= \left( -\frac{v_0}{g}, \frac{v_0 \textrm{cot} \theta_0}{g} \right)$, therefore $(x,y)= \left( -\frac{v_0}{g} + x_0, \frac{v_0 \textrm{cot} \theta_0}{g}+y_0 \right)$ and the radius of the circle is $\frac{v_0}{g \textrm{sin} \theta_0}.$\par
A ray path has been calculated for a velocity gradient with velocity 3000m/s at $x=0$ and 7200m/s at $x=200$, with a source at $(x_0, y_0) = (1, 30)$ and receiver at $(x_1, y_1) = (199, 180)$ (using grid coordinates, the length of a cell is used for scaling after the ray is calculated), which when using grid coordinates gives the parameters $g=21$ and $v_0 = 3021$. From these parameters the true solution is calculated as a section of a circle with center (-142.857, 425.571) and radius of 420.918.
\begin{figure}
\centering
\includegraphics[scale=0.6]{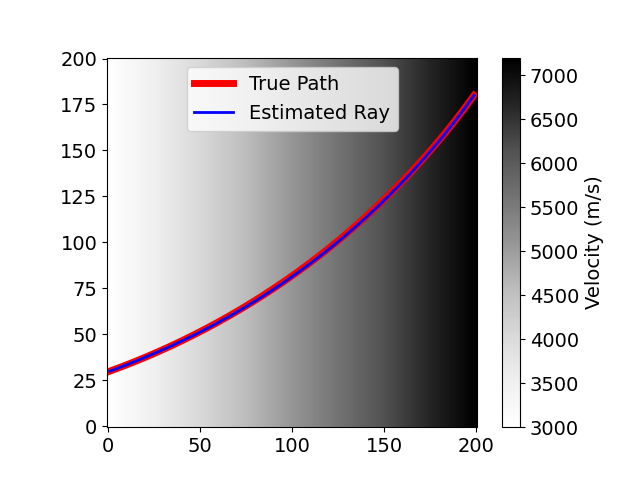}
\caption{Plot of both true and calculated ray paths. Background color represents isotropic velocity at each point in the model.}
\label{fig:ray_vel_grad}
\end{figure}
Figure \ref{fig:ray_vel_grad} shows that the calculated ray path is very close to the true solution. The maximum distance between these lines is approximately 0.3 grid points away from the true solution and true travel times are calculated by integrating along the ray path (using cell width of 0.001) giving travel times using the discretised grid of 5.0884409e-05 for the estimated ray and 5.0884053e-05 for the true solution, giving an error of 0.0007$\%$. The error obtained is dependent on the error in the travel time field and the geometry used. For anisotropic media the error is likely to be higher. From Table \ref{tab:ray_times}, the error for the ray tracing using different plane directions is 0.096{\%} (3 decimal places), on the discretised grid. However if the errors in the travel time field were lower, the errors in ray tracing are expected to be lower.

\subsection{Ray tracing through an anisotropic weld}
\begin{figure}[h]
\centering
\includegraphics[scale=0.3]{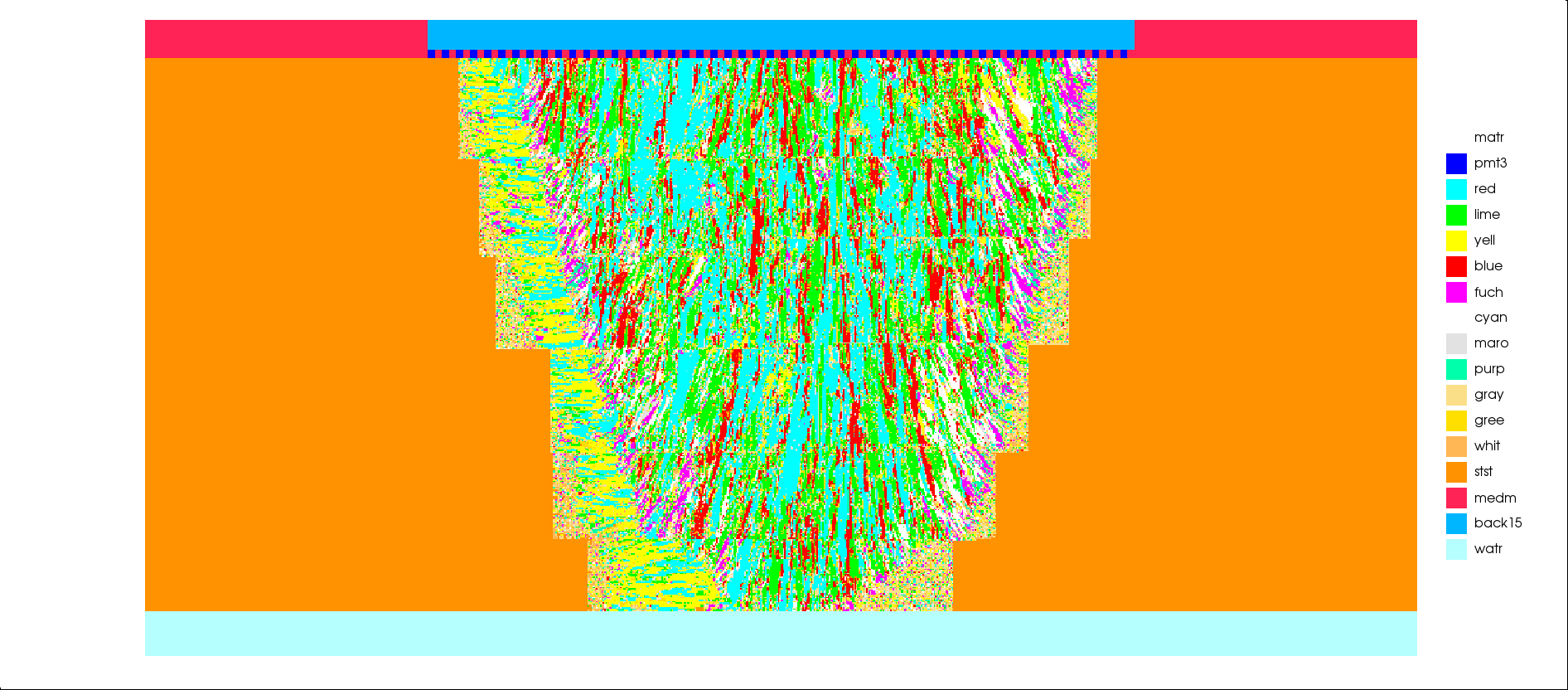}
\caption{Model of a weld (from \cite{anis_weld}). Background colours represent the material at each point (see legend) with materials pmt3 being transducers, medm and back15 are part of the transducer array, watr is water, stst is an isotropic steel and all other materials are different anisotropic orientations of the same austenitic steel which make up the weld structure.}
\label{fig:weld_before}
\end{figure}

\begin{figure}[h]
\centering
\includegraphics[scale=0.4]{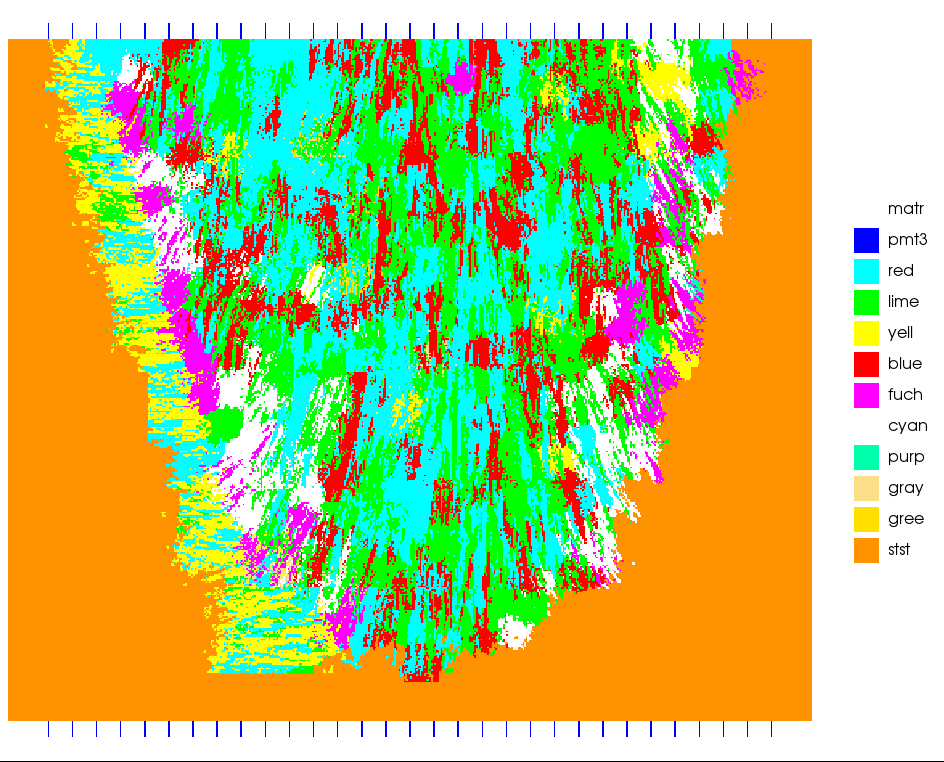}
\caption{Model of a weld after clustering which is used as our model. Background represent the material at each point (see legend), using the same materials as Figure \ref{fig:weld_before}}.
\label{fig:weld_after}
\end{figure}

Ray tracing was performed using a model based on an anisotropic weld obtained using electron back-scatter diffraction \cite{anis_weld} (Figure \ref{fig:weld_before}). The model was simplified using k-means clustering to remove some of the high spatial frequency heterogeneity to create a more reasonable boundary between the weld and parent material. Sources/receivers were also changed so as to act more similarly to a point source and were put on the top and bottom so ray tracing could be applied to the model (see Figures \ref{fig:weld_before} and \ref{fig:weld_after}). The crystal orientations were also changed from a 3D orienation to 2D orientation since we require 2D velocity curves for this version of our codes. Travel times were estimated using A-scans obtained using finite element (FE) software Onscale with the driving function in time shown in Figure \ref{fig:
e_func}. While the driving function is not realistic, it helps to measure more accurate travel time estimates from the modelled waveform due to the rapid onset of energy.

Travel times were estimated by finding the first minimum/maximum above an arbitrarily chosen threshold. A second threshold is then calculated to be the pressure at the min/max divided by 100, and the travel time is estimated as the first time point above this threshold. Due to the method used for estimating these travel times there will be some error in the travel times obtained.

\begin{figure}[h]
\centering
\includegraphics[scale=0.5]{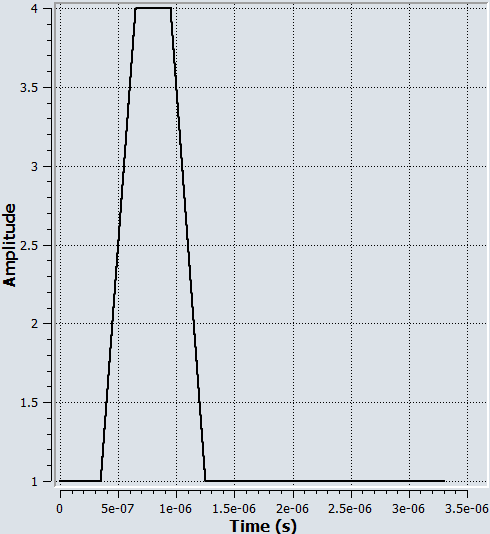}
\caption{Source driving function with time, used for finite element simulation.}
\label{fig:drive_func}
\end{figure}

\begin{figure}
\centering
\includegraphics[width=0.95\textwidth]{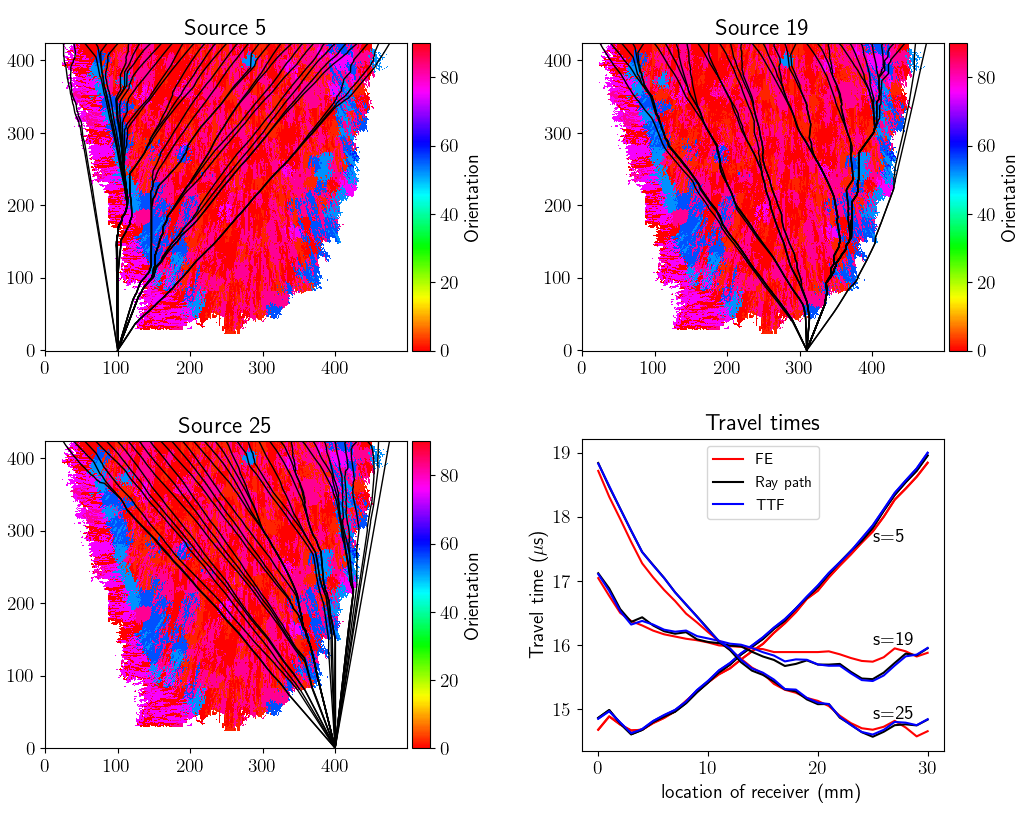}
\caption{(top left),(top right) and (bottom left) are ray paths from sources 5,19 and 25. The background is the anisotropic orientation. The parent material is isotropic so no orientation is given for this material. (bottom right) is a comparison of travel times obtained from ray paths, FE software, and ALI-FMM travel time fields.}
\label{fig:ray_FE_trav_times}
\end{figure}

Comparing travel times calculated along ray paths and those from FE in Figure \ref{fig:ray_FE_trav_times}, we see that differences are relatively small, however for source 19 the differences are larger. The travel times along many ray paths are lower than those from the FE software, and since ray paths always give travel times that are higher than the true solution on a discretised grid, this is a good result.

\section{Discussion}
To improve accuracy within the travel time fields a higher order finite difference approximation could be used by adding an additional point in each stencil to get three points with equal travel times for the estimated wavefront. These three points allow the wavefront to be estimated including its curvature. This would require a different method for the finite difference approximation, and there may be numerous challenges using such methods. If there is any bias towards higher curvature in the wavefront, the curvature in the modelled wavefront may increase and cause unstable and inaccurate results. Another way to reduce errors is to change the density of grid points based on the curvature in the wavefront since the highest errors occur in regions of high curvature. This would have a similar effect as the introduction of finer grids around the source which are used due to high curvature in that area. Since we would only increase the grid density when there is high curvature, this would be computationally more efficient than increasing the grid density for the whole grid. Note that shear wave (S-wave) arrival times could also be computed using the ALI-FMM finite difference calculations by substituting in the shear-vertical or shear-horizontal velocity curves and considering the other eigenvalues/solutions in Section \ref{sect:velocities}. However, since this method was developed with traveltime tomography applications in mind, this study restricted examination to longitudinal waves only since these are the fastest and hence first arriving, meaning they are easier to extract from the collected data.

ALI-FMM can be used in 3D wave propagation using the same techniques used in this paper, however the wavefront must be approximated using a 2D plane. This would require a new set of stencils to approximate the wavefront. Since three points with equal travel times are then required to estimate the wavefront, an additional point is required in the stencil compared to the 2D stencils used in this paper (hence a total of 5 points). The wavefront can also be estimated using curvature with an additional point using the same method proposed above for the 2D case. Ray tracing can be used in 3D by finding where the ray path crosses a set of 2D planes. Due to the additional plane directions that would be used and the number of points within each plane, ray tracing without adaptively changing plane direction would be computationally expensive.

The model used for ray tracing in this paper is input by giving material properties at all points in a grid. When the geometry is put onto a finer grid the detail in the geometry remains unchanged as grid points are assigned the material properties of the closest point in the original grid. If a different method is used that gives a smoother interface between regions, the accuracy of the ray path should improve. For example, in Figure \ref{fig:ray_changing_plane} the geometry in the finer grid can be chosen using the equation of the circle. Modelling the geometry using Voronoi tessellation may be especially useful in NDT applications with polycrystalline materials \cite{Bourne_2021}. Additionally once the ray path has been calculated we can remove points when we know that the ray must travel in a straight line. This would improve accuracy since the fastest ray path between two points in a homogeneous medium is a straight line. Another way to do this would be to look at where the ray path changes material, instead of using an intersection plane during the ray tracing which would be computationally cheaper. For example in Figure \ref{fig:ray_changing_plane}, the ray path should only change direction at the interface between the outside and inside of the circular region, so the ray path could be reduced to four points. These various improvements are left for further work, should they be required.

\section*{Acknowledgements}
This work was funded by the Engineering and Physical Sciences Research Council (EPSRC) (grant numbers EP/S001174/1 and EP/V520032/1).

\section*{Declaration of competing interests}
The authors declare that they have no known competing financial or personal relationships that could have appeared to influence the work reported in this paper.

\section*{Code package}
A code package for running ALI-FMM and ray tracing is provided through a GitHub repository: \url{https://github.com/WiPi-UoS/ALI-FMM-and-ray-tracing}.

\bibliographystyle{unsrt} 
\bibliography{references}

\newpage
\appendix
\section{ALI-FMM Stencils}
Appendix shows plots of the stencils used for ALI-FMM. Figure \ref{fig:apendix_stencils1} gives the square stencils while Figure \ref{fig:apendix_stencils2} gives the triangular stencils.
\begin{figure}[h]
\centering
\begin{subfigure}[b]{0.35\textwidth}
\centering
\includegraphics[width=0.9\textwidth]{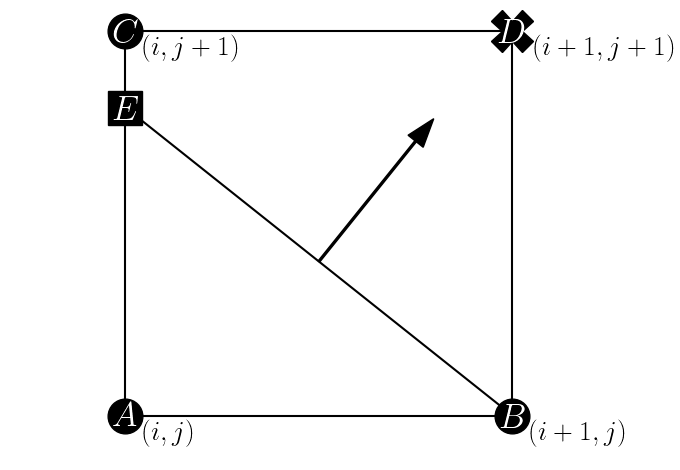}
\caption{Stencil 1-1}
\end{subfigure}
\hspace{0.05\textwidth}
\begin{subfigure}[b]{0.35\textwidth}
\centering
\includegraphics[width=0.9\textwidth]{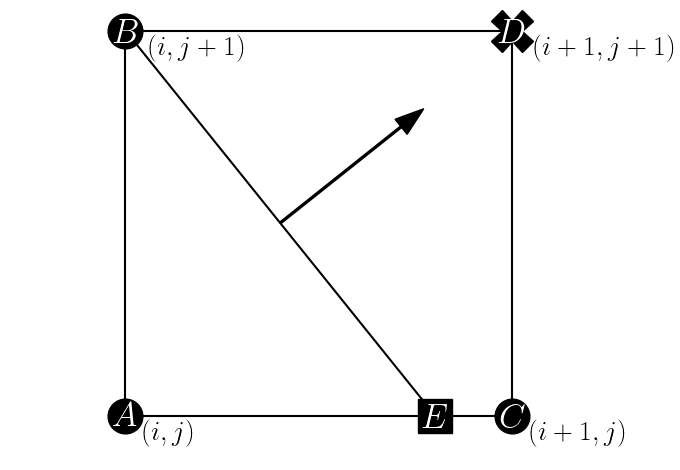}
\caption{Stencil 1-2}
\end{subfigure}
\begin{subfigure}[b]{0.35\textwidth}
\centering
\includegraphics[width=0.9\textwidth]{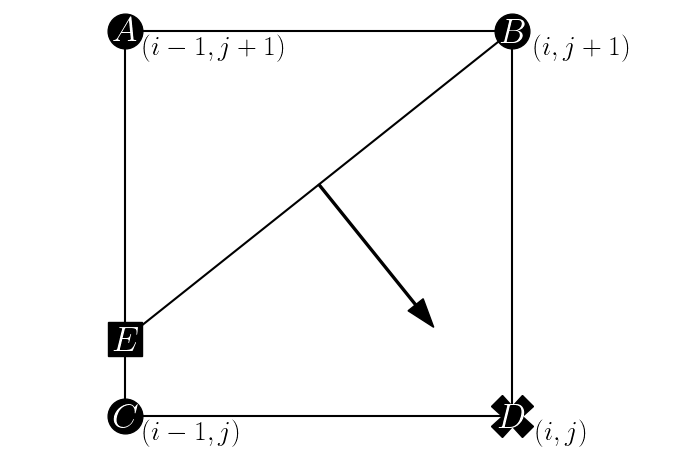}
\caption{Stencil 2-1}
\end{subfigure}
\hspace{0.05\textwidth}
\begin{subfigure}[b]{0.35\textwidth}
\centering
\includegraphics[width=0.9\textwidth]{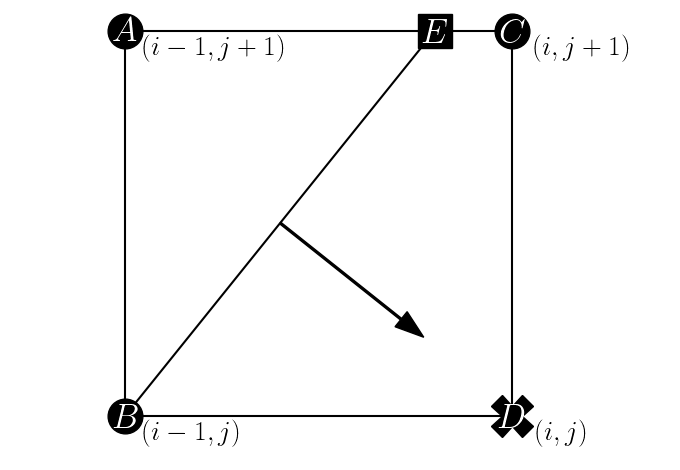}
\caption{Stencil 2-2}
\end{subfigure}
\begin{subfigure}[b]{0.35\textwidth}
\centering
\includegraphics[width=0.9\textwidth]{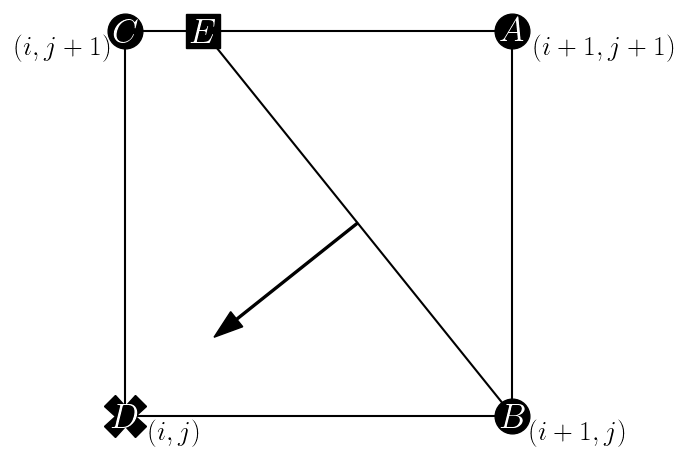}
\caption{Stencil 3-1}
\end{subfigure}
\hspace{0.05\textwidth}
\begin{subfigure}[b]{0.35\textwidth}
\centering
\includegraphics[width=0.9\textwidth]{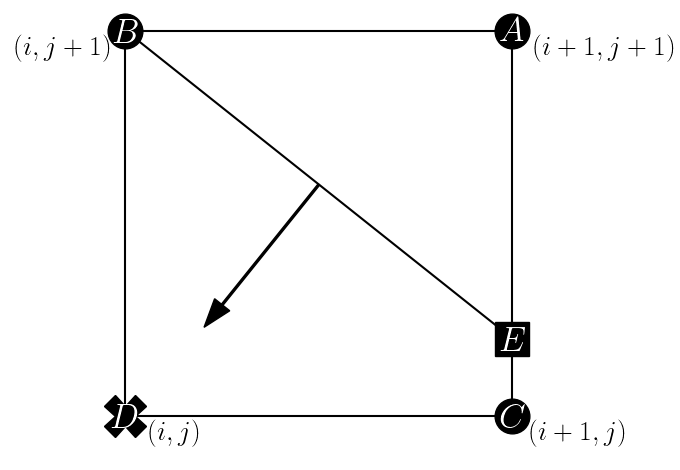}
\caption{Stencil 3-2}
\end{subfigure}
\begin{subfigure}[b]{0.35\textwidth}
\centering
\includegraphics[width=0.9\textwidth]{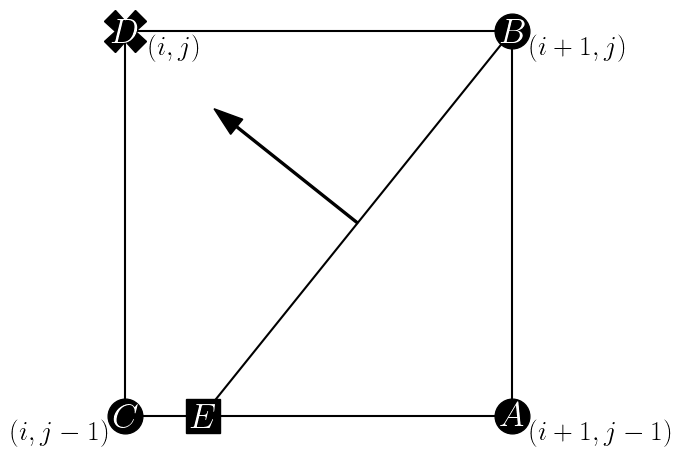}
\caption{Stencil 4-1}
\end{subfigure}
\hspace{0.05\textwidth}
\begin{subfigure}[b]{0.35\textwidth}
\centering
\includegraphics[width=0.9\textwidth]{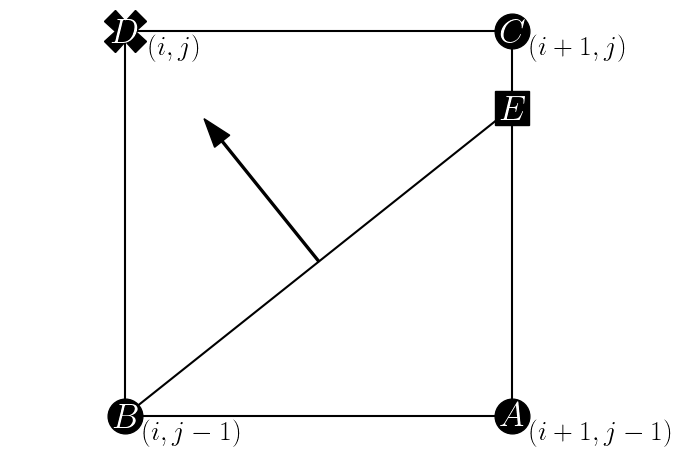}
\caption{Stencil 4-2}
\end{subfigure}
\caption{Square stencils used for finite difference method. Crosses are the points where the travel time is being estimated, the square is a point with unknown location and circles have a known travel time.}
\label{fig:apendix_stencils1}
\end{figure}
\begin{figure}[h]\ContinuedFloat
\centering
\begin{subfigure}[b]{0.45\textwidth}
\includegraphics[width=\textwidth]{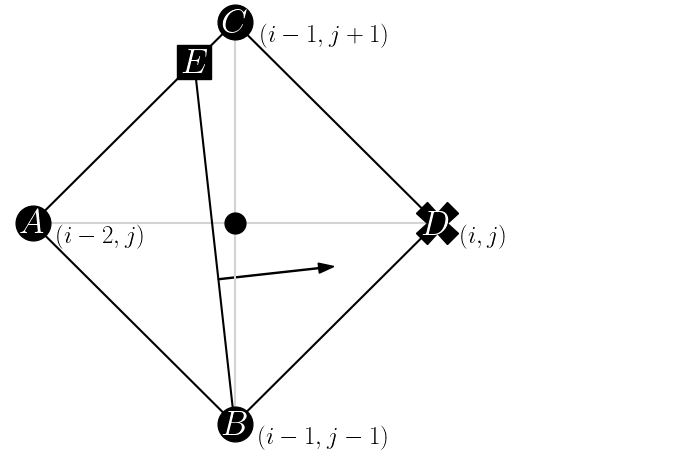}
\caption{Stencil 5-1}
\end{subfigure}
\hspace{0.05\textwidth}
\begin{subfigure}[b]{0.45\textwidth}
\includegraphics[width=\textwidth]{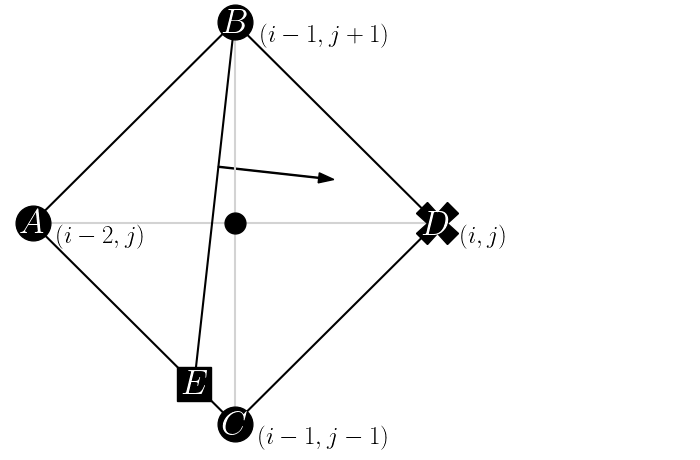}
\caption{Stencil 5-2}
\end{subfigure}
\par\bigskip
\begin{subfigure}[b]{0.45\textwidth}
\includegraphics[width=\textwidth]{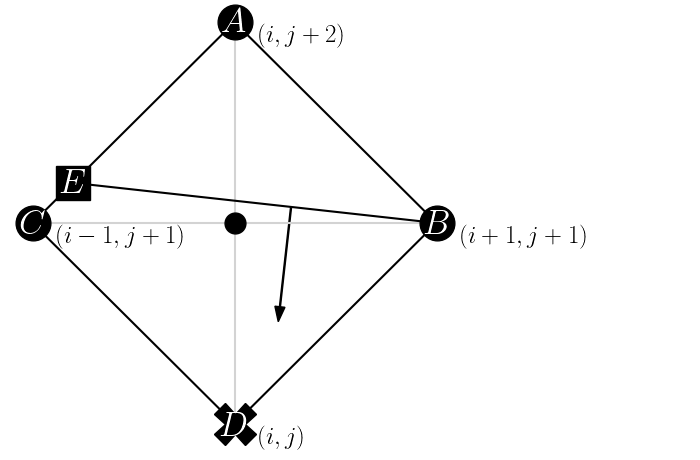}
\caption{Stencil 6-1}
\end{subfigure}
\hspace{0.05\textwidth}
\begin{subfigure}[b]{0.45\textwidth}
\includegraphics[width=\textwidth]{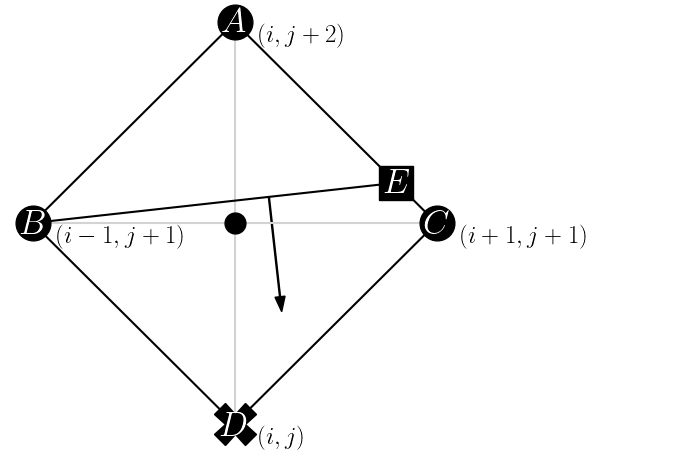}
\caption{Stencil 6-2}
\end{subfigure}
\begin{subfigure}[b]{0.45\textwidth}
\includegraphics[width=\textwidth]{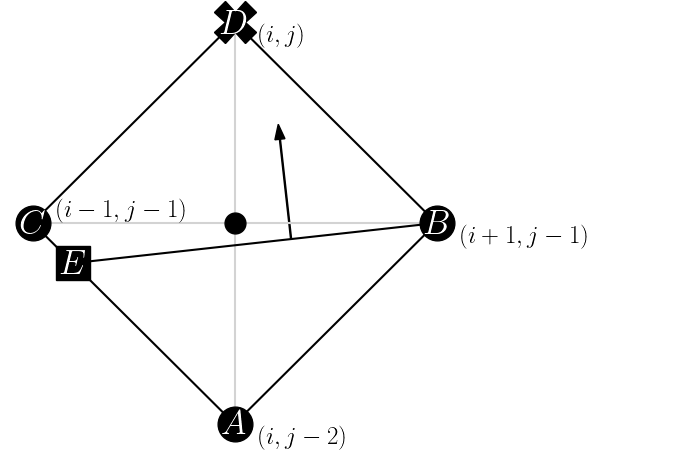}
\caption{Stencil 7-1}
\end{subfigure}
\hspace{0.05\textwidth}
\begin{subfigure}[b]{0.45\textwidth}
\includegraphics[width=\textwidth]{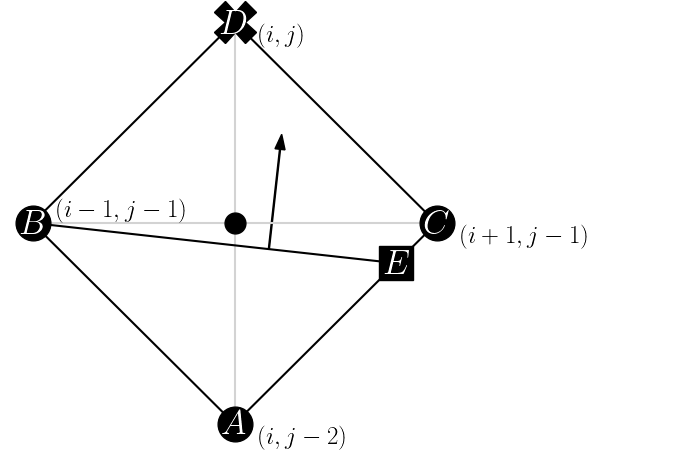}
\caption{Stencil 7-2}
\end{subfigure}
\par\bigskip
\begin{subfigure}[b]{0.45\textwidth}
\includegraphics[width=\textwidth]{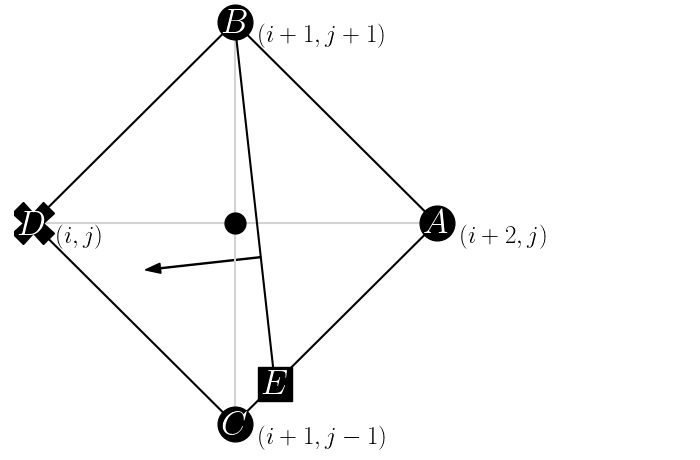}
\caption{Stencil 8-1}
\end{subfigure}
\hspace{0.05\textwidth}
\begin{subfigure}[b]{0.45\textwidth}
\includegraphics[width=\textwidth]{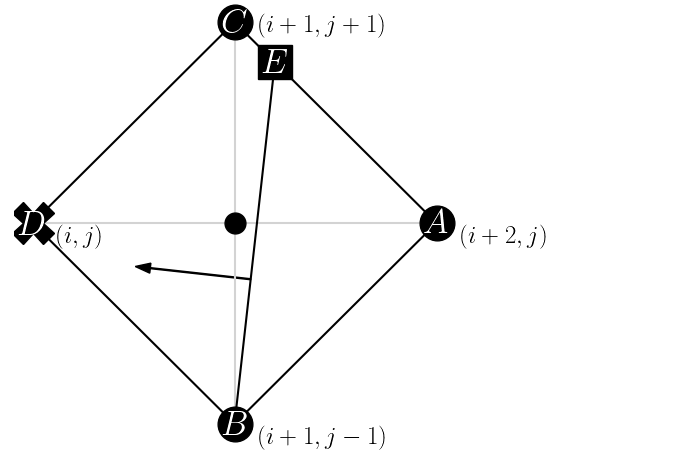}
\caption{Stencil 8-2}
\end{subfigure}
\caption{Square stencils used for finite difference method. Crosses are the points where the travel time is being estimated, the square is a point with unknown location and circles have a known travel time. (cont.)}
\end{figure}

\begin{figure}[h]
\centering
\begin{subfigure}[b]{0.3\textwidth}
\centering
\includegraphics[width=0.8\textwidth]{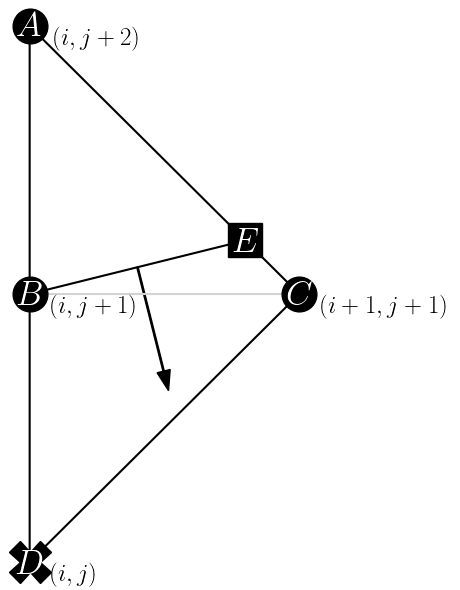}
\caption{Stencil 9-1}
\end{subfigure}
\hspace{0.03\textwidth}
\begin{subfigure}[b]{0.3\textwidth}
\centering
\includegraphics[width=0.8\textwidth]{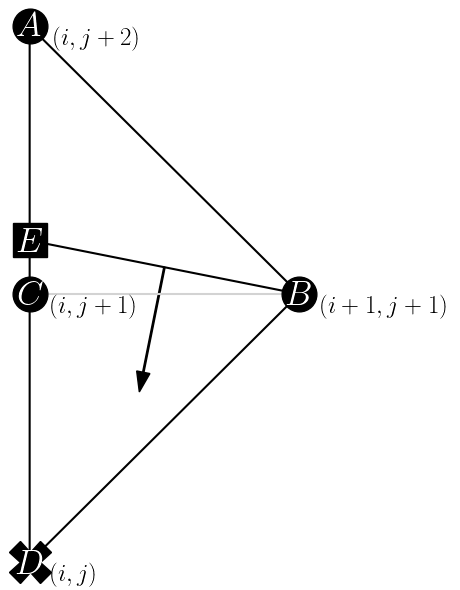}
\caption{Stencil 9-2}
\end{subfigure}
\hspace{0.03\textwidth}
\begin{subfigure}[b]{0.3\textwidth}
\centering
\includegraphics[width=0.8\textwidth]{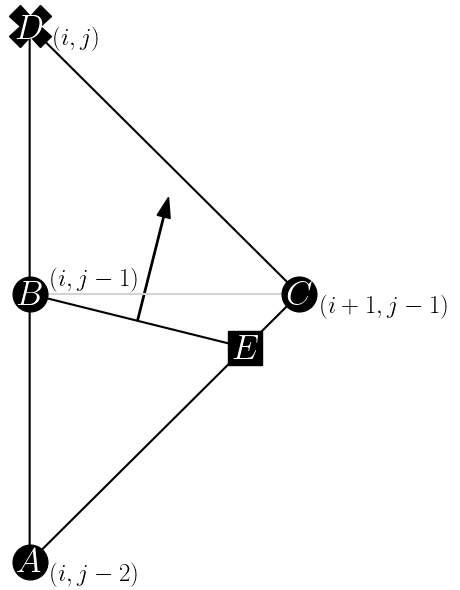}
\caption{Stencil 10-1}
\end{subfigure}
\par
\begin{subfigure}[b]{0.3\textwidth}
\centering
\includegraphics[width=0.8\textwidth]{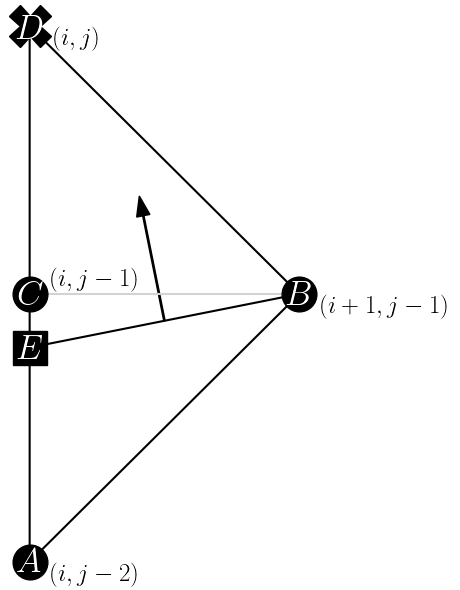}
\caption{Stencil 10-2}
\end{subfigure}
\hspace{0.03\textwidth}
\begin{subfigure}[b]{0.3\textwidth}
\centering
\includegraphics[width=0.8\textwidth]{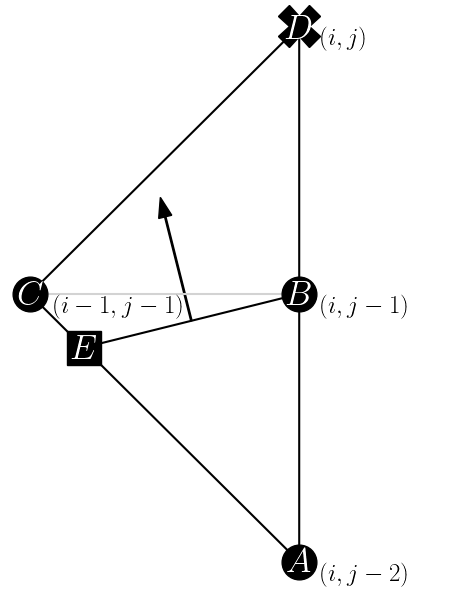}
\caption{Stencil 11-1}
\end{subfigure}
\hspace{0.03\textwidth}
\begin{subfigure}[b]{0.3\textwidth}
\centering
\includegraphics[width=0.8\textwidth]{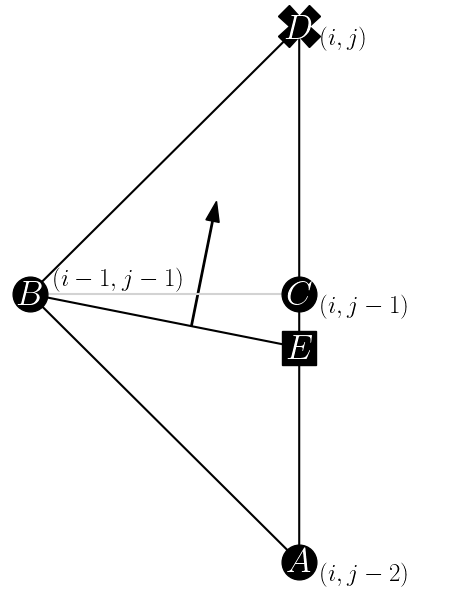}
\caption{Stencil 11-2}
\end{subfigure}
\begin{subfigure}[b]{0.3\textwidth}
\centering
\includegraphics[width=0.8\textwidth]{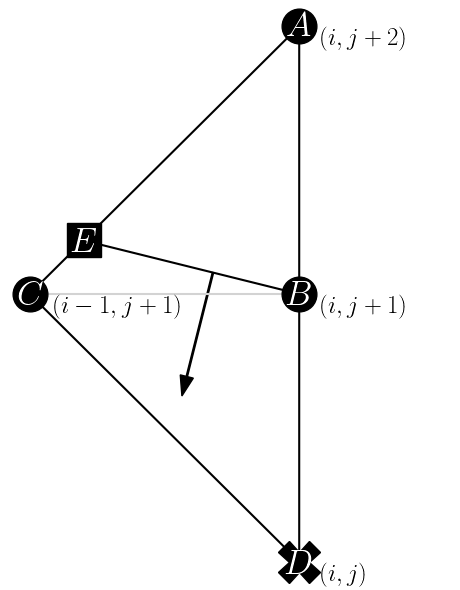}
\caption{Stencil 12-1}
\end{subfigure}
\hspace{0.03\textwidth}
\begin{subfigure}[b]{0.3\textwidth}
\centering
\includegraphics[width=0.8\textwidth]{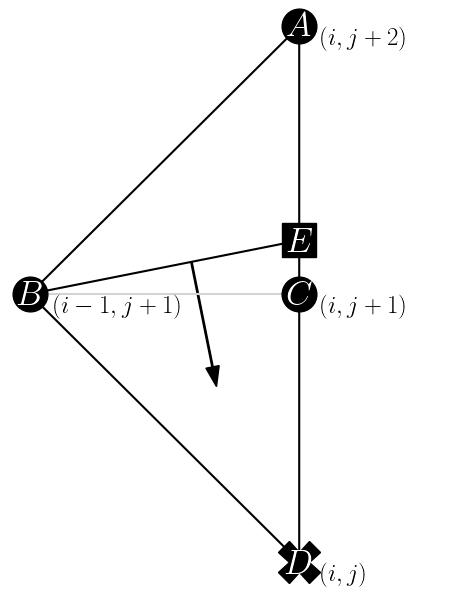}
\caption{Stencil 12-2}
\end{subfigure}
\caption{Triangular stencils used for finite difference method. Crosses are the points where the travel time is being estimated, the square is a point with unknown location and circles have a known travel time.}
\label{fig:apendix_stencils2}
\end{figure}

\begin{figure}[h]\ContinuedFloat
\centering
\begin{subfigure}[b]{0.45\textwidth}
\includegraphics[width=\textwidth]{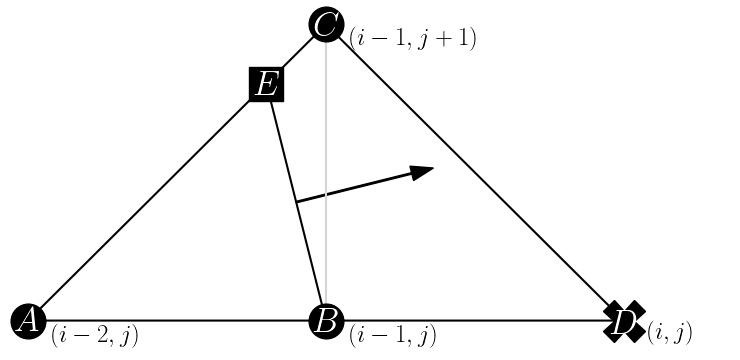}
\caption{Stencil 13-1}
\end{subfigure}
\hspace{0.05\textwidth}
\begin{subfigure}[b]{0.45\textwidth}
\includegraphics[width=\textwidth]{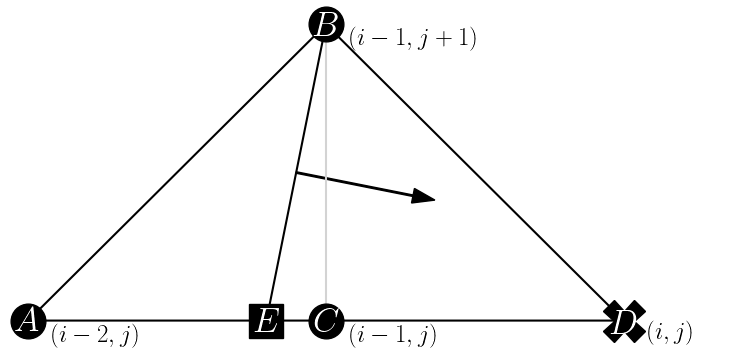}
\caption{Stencil 13-2}
\end{subfigure}
\par\bigskip
\begin{subfigure}[b]{0.45\textwidth}
\includegraphics[width=\textwidth]{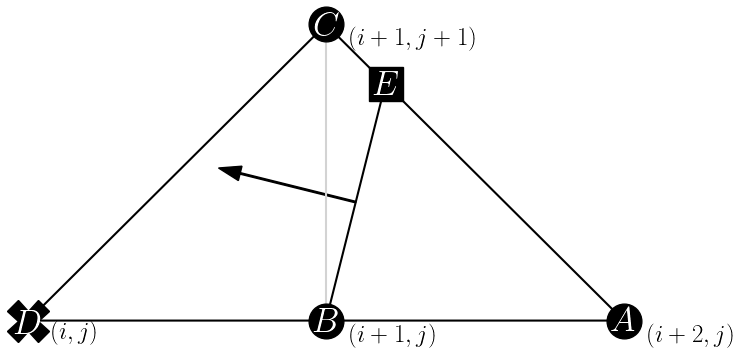}
\caption{Stencil 14-1}
\end{subfigure}
\hspace{0.05\textwidth}
\begin{subfigure}[b]{0.45\textwidth}
\includegraphics[width=\textwidth]{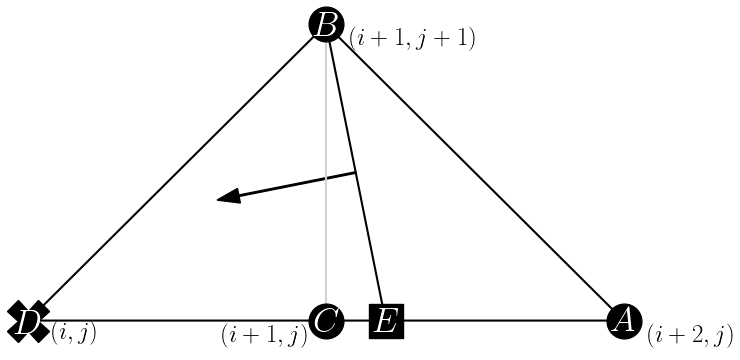}
\caption{Stencil 14-2}
\end{subfigure}
\par\bigskip
\begin{subfigure}[b]{0.45\textwidth}
\includegraphics[width=\textwidth]{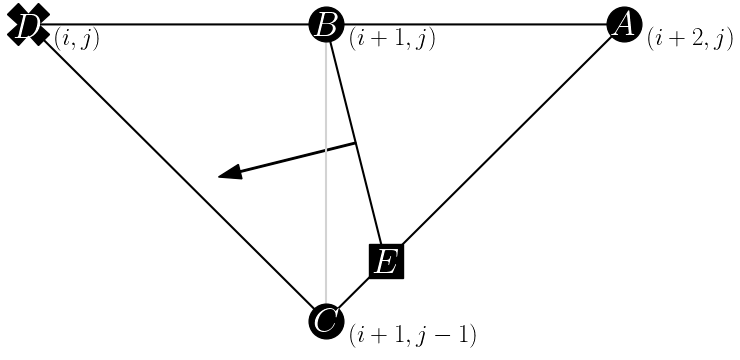}
\caption{Stencil 15-1}
\end{subfigure}
\hspace{0.05\textwidth}
\begin{subfigure}[b]{0.45\textwidth}
\includegraphics[width=\textwidth]{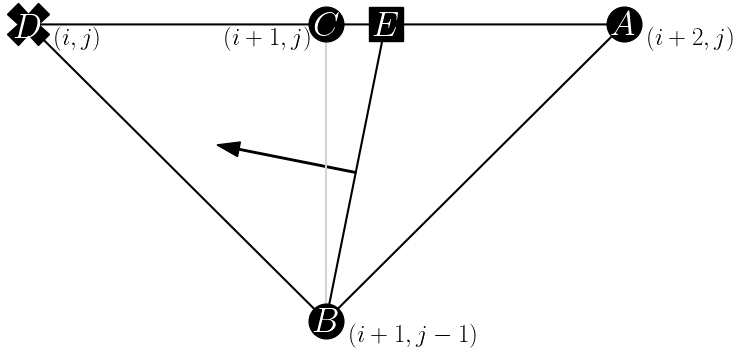}
\caption{Stencil 15-2}
\end{subfigure}
\par\bigskip
\begin{subfigure}[b]{0.45\textwidth}
\includegraphics[width=\textwidth]{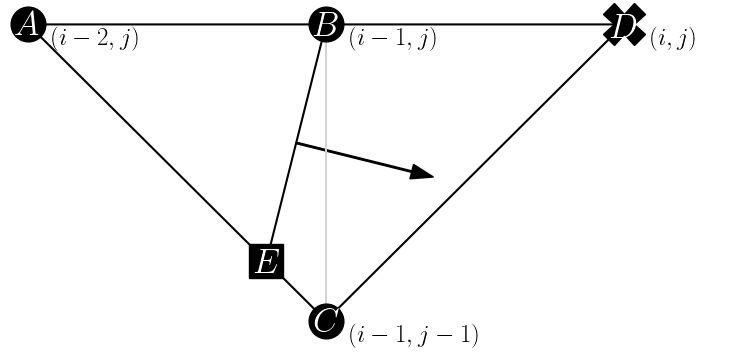}
\caption{Stencil 16-1}
\end{subfigure}
\hspace{0.05\textwidth}
\begin{subfigure}[b]{0.45\textwidth}
\includegraphics[width=\textwidth]{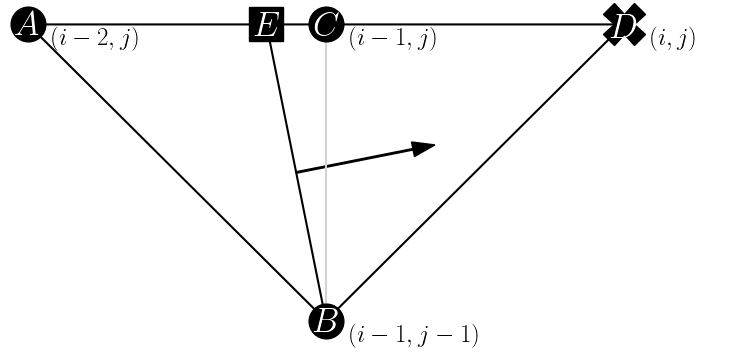}
\caption{Stencil 16-2}
\end{subfigure}
\caption{Triangular stencils used for finite difference method. Crosses are the points where the travel time is being estimated, the square is a point with unknown location and circles have a known travel time. (cont.)}
\end{figure}

\end{document}